\documentclass{aa}  

\usepackage{graphicx}
\usepackage{float}
\usepackage{subfig}

\usepackage{txfonts}
\begin{document}

   \title{The power of wavelets in analysis of transit and phase curves in presence of stellar variability and instrumental noise}

   \subtitle{I. Method and validation}

   \author{Sz. Csizmadia
          \inst{1}
          \and
          A.M.S. Smith\inst{1}
          \and
          J. Cabrera\inst{1}
          \and
          P. Klagyivik\inst{1}
          \and
          A. Chaushev\inst{2}
          \and
          K. W. F. Lam\inst{3}
          }

   \institute{Deutsches Zentrum f\"ur Luft- und Raumfahrt, Institute of Planetary Research,
              Rutherfordtstrasse 2, D-12489 Berlin, Germany\\
              \email{szilard.csizmadia@dlr.de}
              \and
              Department of Physics and Astronomy, University of California,
              Irvine, 4129 Frederick Reines Hall, Irvine, CA, USA, 92697
              \and
              Center for Astronomy and Astrophysics, Technical University, Berlin, Hardenbergstr. 36, 10623 Berlin, Germany
             }

   \date{Received May 13, 2021; accepted Month Day, 2021}

 
  \abstract
   {Stellar photometric variability and instrumental effects, like cosmic ray hits, data discontinuities, data leaks, instrument aging etc. cause difficulties in the characterization of exoplanets and have an impact on the accuracy and precision of the modelling and detectability of transits, occultations and phase curves.}
   {This paper aims to make an attempt to improve the transit, occultation and phase-curve modelling in the presence of strong stellar variability and instrumental noise. We invoke the wavelet-formulation to reach this goal.}
   {We explore the capabilities of the software package Transit and Light Curve Modeller (TLCM). It is able to perform a joint radial velocity and light curve fit or light curve fit only. It models the transit, occultation, beaming, ellipsoidal and reflection effects in the light curves (including the gravity darkening effect, too). The red-noise, the stellar variability and instrumental effects are modelled via wavelets. The wavelet-fit is constrained by prescribing that the final white noise level must be equal to the average of the uncertainties of the photometric data points. This helps to avoid the overfit and regularizes the noise model. The approach was tested by injecting synthetic light curves into Kepler's short cadence data and then modelling them.}
  {The method performs well over a certain signal-to-noise (S/N) ratio. 
  We give limits in terms of signal-to-noise ratio for every studied system parameter which is needed to accurate parameter retrieval. The wavelet-approach is able to manage and to remove the impacts of data discontinuities, cosmic ray events, long-term stellar variability and instrument ageing, short term stellar variability and pulsation and flares among others.}
   {We conclude that precise light curve models combined with the wavelet-method and with well prescribed constraints on the white noise are able to retrieve the planetary system parameters, even when strong stellar variability and instrumental noise including data discontinuities are present.}

   \titlerunning{Power of wavelets in light curve analysis}
   \authorrunning{Csizmadia Sz. et al.}

   \keywords{
     Methods: data analysis -- Planets and satellites: atmospheres -- Planets and satellites: interiors -- Planets and satellites: general -- Techniques: photometric}

   \maketitle

%
\section{Introduction}\label{sec:intro}

The light curve of an exoplanetary system may show transits, occultations and phase-curve variations. The transit technique offers a unique opportunity to determine the accurate radii of transiting exoplanets. Complementing the photometric transit observations with radial velocity data, the planetary mass and mean density can be determined. The phase-curves describe the scattering and reflecting properties of an atmosphere at different orbital phase. Phase-curves and occultations are considered as the best opportunity to study the three-dimensional structure of planetary atmospheres \citep{parmentier17,winn10}. 

In such transit- as well as in phase curve-analysis, the following four problems can arise. (i) The stellar activity, stellar variability -- including pulsation and granulation, too -- and instrumental noise cause difficulties to find and to restore the exact shape of the transit- and phase-curves (e.g. \citealt{oshagh18,sulis20}). The transit depth can be also affected by stellar spots yielding wrong planetary radii. If sudden and discontinouos flux variations (jumps) occur in the flux measurements due to a cosmic ray hit or other kind of instrumental effect, then there is a difficulty to establish the mean flux level of the host star. This can lead to further change in the transit depth because the normalized flux level is different and maybe not well fitted to each other before and after the jump. (ii) The beaming-effect might be degenerate with the reflection effect and care is needed to separate them from each other \citep{csizmadia2020}. (iii) The ellipsoidal effect, when it is significant, must be also separated from the phase-variation. The ellipsoidal and the reflection effects have higher order harmonics of the orbital frequency and if it is not modelled carefully, the badly modelled ellipsoidal effect  can affect the shape of the reflection curve and this may lead to misconclusions. (iv) In addition, the exact shape of the phase curve is not known without a detailed {\it a priori} knowledge of the atmosphere (composition, scattering and reflecting properties, scale height, clouds, particle sizes of the aerosols etc, \citealt{garcia15}).

We make extensive tests on synthetic light curves to overcome problem (i), namely the stellar and instrumental noise sources are modelled by a wavelet-transform. We investigated how well the wavelet-method can filter out the stellar variability and instrumental noise effects. We show in the present study that the wavelet-transform is a powerful tool to model flux-variations of stellar and instrumental origin which increase the accuracy and precision of parameter retrieval. In a subsequent paper we apply our method to KELT-9b (Csizmadia et al, 2021,  submitted, Paper II).

To solve problems (ii-iv) one way can be to use prescribed forms of phase-curves and to improve the description of the ellipsoidal effect. In Paper II we attempt to fit single cosinusoidal, Lambertian, Kane-Gelino- and Kopal-type phase curves to the time-series data of KELT-9b obtained by the TESS space telescope. These four different phase functions yield significantly different shaped reflection curves. In Section~\ref{sec:phase_curve} we detail the model of the ellipsoidal, beaming and reflection effects used for the fine-analysis of the KELT-9b light curve in Paper II. We also update the gravity darkening model of TLCM in Section~\ref{sec:gravity_darkening}. The wavelet model and its test are presented on Section~\ref{sec:wavelet_method}. The summary of this study and our conclusions can be found in Section~\ref{sec:summary}.

\section{Model of out-of transits variation}
\label{sec:phase_curve}

The out-of-transits variation is usually divided into components of reflection and ellipsoidal effects. When it became observable with space-based telescopes, this list was extended by the beaming-effect component \citep{zucker07,faigler11}.

the sum of the ellipsoidal, beaming and reflection (phase-curve) effects. We describe our model of these effects hereafter. There are other kind of photometric variations which we consider part of our red noise model (see Section~\ref{sec:wavelet_method}): stellar pulsation, stellar activity, possible additional eclipses caused by another star, instrumental and other -- non-white noise-like -- effects etc. Note that we distinguish between phase curve (only the reflection effect, without the beaming and ellipsoidal effects) and the phase function which contains the time-dependence of the phase curve.

\subsection{Phase curve}
\label{subsec:phase_curve}

We utilize Transit and Light Curve Modeller (TLCM, \citealt{csizmadia2020}) which expresses the phase curve variation ($F_\mathrm{ph}$) in the following form:
\begin{equation}
    \frac{F_\mathrm{ph}}{F_\mathrm{star}} =  \left( \frac{I_\mathrm{planet}}{I_\mathrm{star}} \left( \frac{R_\mathrm{planet}}{R_\mathrm{star}} \right)^2 + A_\mathrm{geometric} \left( \frac{R_\mathrm{planet}}{R_\mathrm{star}} \frac{R_\mathrm{star}}{d} \right)^2 \Phi(\alpha) \right) \label{eq:tlcm_phase_function}
\end{equation}
Here $F_\mathrm{ph}$ and $F_\mathrm{star}$ are the reflected and stellar fluxes, respectively. The phase-angle $\alpha$ in the phase function $\Phi$ is 
\begin{equation}
   \cos (\alpha + \varepsilon) = \cos(\omega + v) \sin i_p  
   \label{eq:phase_angle_definition}
\end{equation}
where $\omega$ is the argument of the periastron of the planetary orbit and $v$ is the true anomaly. $i_p$ is the inclination of the planetary orbit. $I$ is the surface specific intensity in the passband of the observation, $d$ is the mutual star-planet actual distance. $R$ denotes the radius. $A_\mathrm{geometric}$ is the so-called wavelength-dependent geometric albedo.

The angle $\varepsilon$ takes into account that there can be a phase-shift in the phase curve due to atmospheric circulation, i.e. the brightest point of the planet can be shifted eastward or westward relative to the substellar point\footnote{Positive values of $\varepsilon$ mean eastward, negative values mean westward shift 
.}
\citep{parmentier17}. The observed values of $\varepsilon$ vary  between wide ranges, from -70 to +50 degrees (\citealt{parmentier16,bell21}) and references therein.

For the star the phase curve can be obtained in a similar way by interchanging the indices appropriately and shifting the phase curve by $180^\circ$. This may be important for detached eclipsing binary stars where TLCM can also be used for modelling and takes all these effects of the two stellar component into account  \citep{csizmadia2020}.

\subsection{Time dependence of the phase function}

The exact form of the phase-function $\Phi$ strongly depends on wavelength, the chemical composition of the atmosphere, the particle size in it, optical depth, clouds properties, single scattering albedo \citep[e.g.][]{garcia15}. Establishing the exact form of the phase function needs a priori knowledge on the atmospheric properties which are not always available. In addition, it requires complex and lengthy numerical calculations \citep{garcia15}. For data analysis, one can try analytically expressed approximate formulae, too. In this paper series we probe the following four phase functions offered by TLCM on KELT-9b as follows.

First one is a simple cosine-like:
\begin{equation}
    \Phi(\alpha) = \frac{1}{2} (1 - \cos \alpha) \label{eq:simple_cosine}
\end{equation}
The second one is a Lambertian:
\begin{equation}
    \Phi(\alpha) = \frac{\sin \alpha - \alpha \cdot \cos \alpha}{\pi} \label{eq:lambertian}
\end{equation}
The third one is taken from \citet{kanegelino2011} where the phase angle must be measured in degrees:
\begin{equation}
    \Phi(\alpha) = 10.0^{-0.4 \cdot (0.09 (\alpha/100^\circ)) + 2.39 (\alpha/100^\circ)^2) - 0.65  (\alpha/100^\circ)^3)} \label{eq:kanegelino}
\end{equation}
This formula is based on the observations of Venus and Jupiter and it takes into account that these planets have significant backward-scattering due to their clouds \citep{hilton1992}. On eccentric orbits the particle properties can change in the atmosphere due to the variable insolation. According to \citet{kanegelino2011}, the geometric albedo in Eq.~(\ref{eq:tlcm_phase_function}) must be replaced by $A' (d)$ for this case as
\begin{equation}
    A' (d) = \left( A_{\mathrm{geometric}} + 0.2 \frac{e^{d-1} - e^{1-d}}{e^{d-1} + e^{1-d}} \right)
    \label{eq:kane_gelino_correction}
\end{equation}
Of course, one can ask how well this Kane-Gelino phase function can perform on hot Jupiters where the response of the atmosphere can be quite different than in the case of the cooler Jupiter and the terrestrial like venusian atmopshere. We try this phase function on the hot Jupiter  KELT-9b in Paper II, too, and this will give an answer. 

The fourth and the last phase function tried here is based on the theory of binary star phase function which takes umbral and penumbral effects into account up to the fourth order of the phase-angle. This is taken from \citet{kopal_1959}:
\begin{equation}
     \Phi(\alpha) = C_0 + C_1 \cos \alpha + C_2 \cos 2\alpha + C_4 \cos 4\alpha \label{eq:kopal}
\end{equation}
where
\begin{equation}
C_0 = \frac{8}{3\pi^2} \left( \frac{R_j}{d} \right)^2 + \frac{1}{16} \left( \frac{R_j}{d} \right)^3 + \frac{2}{3\pi^2} \left( \frac{R_j}{d} \right)^4 - \frac{2 K_j}{\pi^2} \frac{R_j^2 R_{3-j}^2}{d^4} \label{eq:kopal_coeff_0}
\end{equation}
\begin{equation}
C_1 = \frac{1}{3} \left( \frac{R_j}{d} \right)^2 + \frac{1}{4} \left( \frac{R_j}{d} \right)^4 \label{eq:kopal_coeff_1}
\end{equation}
\begin{equation}
C_2 = \frac{16}{27\pi^2} \left( \frac{R_j}{d} \right)^2 + \frac{3}{16} \left( \frac{R_j}{d} \right)^3 + \frac{4}{15\pi^2} \left( \frac{R_j}{d} \right)^4 + \frac{4 K_j}{3\pi^2} \frac{R_j^2 R_{3-j}^2}{d^4} \label{eq:kopal_coeff_2}
\end{equation}
\begin{equation}
C_4 = \frac{16}{675\pi^2} \left( \frac{R_j}{d} \right)^2 - \frac{52}{105\pi^2} \left( \frac{R_j}{d} \right)^4 + \frac{4 K_j}{15\pi^2} \frac{R_j^2 R_{3-j}^2}{d^4} \label{eq:kopal_coeff_4}
\end{equation}
and the limb darkening correction is given for the reflection in a linear form as
\begin{equation}
  K_j = 1 - \frac{12}{5 \pi} \frac{5+(\pi-5) u_1}{3-u_1}  \label{eq:kopal_coeff_limb_darkening}
\end{equation}
$C_3$ is zero according to Kopal (1959). For the planet, one can take linear limb darkening coefficient as $u_1 = 0$ as first order approximation. When one calculate the planet's reflection effect, $j=2$ where $R_2 = R_\mathrm{planet}$; while if one calculates the star's reflection effect $j=1$ with $R_1 = R_\mathrm{star}$. (In case of a detached binary system, these are the primary ($j=1$) and the secondary stars $j=2$, respectively.) This phase function also takes back-warming effects into account which are negligible in star-planet system but it can be important for detached or even closer binary star system.

The planet-to-star radius ratio ($R_\mathrm{planet}/R_\mathrm{star}$) and the scaled semi-major axis ratio ($a/R_\mathrm{star}$)  are known from the transit and light curve analysis or - as in this study - fitted simultaneously with the phase-curve parameters.

In eccentric orbits the star-planet distance $d$ varies as
\begin{equation}
    d = \frac{a(1-e^2)}{1 + \cos v}  \label{eq:star_planet_distance}
\end{equation}
where $e$ is eccentricity (not confuse with the Euler-number in Eq.~\ref{eq:kane_gelino_correction}). 

\subsection{Dayside and nightside emission}
\label{sec:dayside_nightside}

Clearly, the dayside and the nightside phase curves are varying in the opposite phase and therefore we have
\begin{equation}
\Phi_\mathrm{nightside} (\alpha) = 1 - \Phi_\mathrm{dayside} (\alpha)
\end{equation}
The total planetary phase-curve is the sum of the dayside and the nightside emission at a certain phase:
\begin{equation}
    F_\mathrm{ph} = F_\mathrm{nightside} (1-\Phi (\alpha) ) + F_\mathrm{dayside} \left( \frac{a}{d}\right)^2 \Phi (\alpha) \label{eq:dayside_nightside_emission}
\end{equation}
Comparing Eq.~(\ref{eq:dayside_nightside_emission}) to Eq.~(\ref{eq:tlcm_phase_function}) we can relate to the measured quantities to the parameters we search for via:
\begin{equation}
    \frac{F_\mathrm{nightside}}{F_\mathrm{star}} = \frac{I_\mathrm{planet}}{I_\mathrm{star}} \left( \frac{R_\mathrm{planet}}{R_\mathrm{star}} \right)^2 \label{eq:nightside_tlcm}
\end{equation}
and
\begin{equation}
    \frac{F_\mathrm{dayside}}{F_\mathrm{star}} = \frac{F_\mathrm{nightside}}{F_\mathrm{star}} + A_{\mathrm{geometric}} \left( \frac{R_\mathrm{planet}}{R_\mathrm{star}}  \frac{R_\mathrm{star}}{a} \right)^2 \label{eq:dayside_tlcm}
\end{equation}
Note that $I_\mathrm{planet}/I_\mathrm{star}$, $R_\mathrm{planet}/{R_\mathrm{star}}$, $A_{\mathrm{geometric}}$ and reciproc of $R_\mathrm{star}/a$ are fitting parameters in TLCM and they can be measured from the simultaneous fit of transit, occultation and from phase-curve.

We note that we assume the very same phase function for the reflected light and for the dayside/nightside emission in TLCM, which is, of course, just an approximation of reality. However, we use as simple approach as possible.

\subsection{Ellipsoidal effect}

The flux-variation caused by the ellipsoidal shapes of the components is characterized following Kopal (1959) ($j=1$ for the star and $j=2$ for the planet):
\begin{eqnarray}
    \frac{F_\mathrm{ellipsoidal,j}}{F_\mathrm{star}} & = & \nonumber \\  f_j  \frac{M_{3-j}}{M_j} \Sigma_{l=2}^4 w_{jl} \left( \frac{1+e \cos v}{1-e^2} \right)^{l} \left( \frac{R_j}{a} \right)^{l+2} P_{l} \left( cos(v+\omega) \sin i_p \right) \nonumber \\ \label{eq:ellipsoidal_effect}
\end{eqnarray}
with the gravity darkening correction is
\begin{equation}
   \tau = \frac{4 \times 14388.0 \mu m \cdot K}{\lambda T_{eff,j} \left(1.0-e^{-14388.0/\lambda/T_{eff,j}} \right)}
\end{equation}
Here $hc/k_B = 14 388 \mu m \cdot K$, effective wavelength of the observation $\lambda$ is given in microns, $T_\mathrm{eff}$ is in Kelvins, and according to \cite{kopal_1959}:
\begin{equation}
   w_2 =  \frac{2(15 + u_1) \times (1+k_2)}{5(6 - 2u_1 - 3u_2) \left(1+\frac{\tau}{4} \left( \frac{5}{1+k_2}-1 \right) \right)}  
\end{equation}
\begin{equation}
   w_3 = \frac{(35 u_1 + 48 u_2) \times (1.0+k_3)}{7(6 - 2u_1 - 3u_2) \left(1.0+\frac{\tau}{10} \left( \frac{7}{1+k_3}-2.0 \right) \right)}
\end{equation}
\begin{equation}
   w_4 = \frac{9}{8}  \frac{9(4 u_1 + 7u_2 - 4) \times (1.0+k_4)}{8(6 - 2u_1 - 3u_2) \times \left(1.0+\frac{\tau}{18} \left( \frac{9.0}{1.0+k_4}-3.0 \right) \right)}
\end{equation}
Limb darkening coefficients $u_{1,2}$ of \citet{kopal_1959} are from the transit fit. They are related to the limb darkening coefficients $u_a$ and $u_b$ of \cite{claret04} via $u_1 = u_a + 2\cdot u_b$ and $u_2 = -u_b$. (Of course, the limb darkening coefficients can be different for the two objects in the system.) The apsidal motion constants $k_i$ are functions of stellar mass, metalicity, radius and evolutionary status. The $k_i$s of the primary star or of the host star are fixed at their theoretically calculated values of Claret (2004). Note that the apsidal motion constant is half of the Love-number \citep{csizmadia19}. $f_1 = 1$ for the star and $f_2 = I_{planet} / I_{star} (R_\mathrm{planet}/R_\mathrm{star})^2$ for the planet (or secondary star) in the respective photometric passband (same as in Eq.~\ref{eq:nightside_tlcm}).

\subsection{Beaming effect}

The beaming effect is characterized as
\begin{equation}
    B_j = \pm f_j \left(\frac{R_j}{a} \right)^2 \Omega_j \left(T_{eff,j}, \log g_j, Z_j\right) \frac{K_j}{c} (e \cos \omega + cos(v(t) + \omega))
\end{equation}
Here $B$ is the contribution of the beaming effect to the observed flux in units of stellar-flux, $\Omega_j$ is the spectral index derived from theoretical stellar spectra of \cite{munari05}, effective temperature, surface gravity $\log g$, and metalicity $Z$. These spectra were convolved with the response function of used photometer \citep{csizmadia2020}. The star and the companion has the index $j$ as before. $K$ and $c$ are the radial velocity amplitudes and speed of the light, respectively. The plus sign is valid for the companion and the minus sign stands for the star.

For further details on the reflection, ellipsoidal and beaming  effects see \cite{csizmadia2020}.

\section{Gravity darkening} \label{sec:gravity_darkening}

TLCM is able to model the gravity darkening effect with some simplifications and therefore our treatment is valid only for planets where $R_{planet} / R_{star} < 0.2$. Our approach was first presented as part of the study in \cite{lendl20}. Therefore we give only the following details.

Gravity darkening was invoked via a semi-small-planet approximation: the limb darkening and the transit event were calculated via the precise, analytic formulae of \citet{mandel2002}. The effect of gravity darkening was taken into account in the following way. The local surface effective temperature was calculated from 
\begin{equation}
    T_{local} = T_{\ast} \left(\frac{|\nabla V|)}{|\nabla V|_{pole}} \right)^{\beta} \label{eq:gravity_darkening}
\end{equation}
where the local surface potential\footnote{The stellar gravitational potential $V = GM_{star}/R_{star}$ was expressed by easier measurable quantities via Kepler's third law.} is with $q = M_{planet} / M_{star}$:
\begin{equation}
    V = \frac{n^2 a^3}{(1+q) r} + \frac{1}{2} \omega_{rot}^2 r^2 \sin^2 b
\end{equation}
%
Te polar temperature is not equal to the mean effective temperature of the star in the case of rotating stars (cf. Eq. 8 of \citealt{wilson79}). The mean motion is denoted by $n$ and $b$ the astrographic latitude. The rotational angular velocity $\omega_\mathrm{rot}$ can be calculated from the known stellar radius, the measured or fitted $V_{rot} \sin I_{star}$, and the fitted stellar inclination:
\begin{equation}
    \omega_\mathrm{rot} = \frac{(V\sin I_\mathrm{star})_{sp}}{ R_\mathrm{star} \sin I_\mathrm{star}}
\end{equation}
The spectroscopically measured $(V \sin I_\mathrm{star})_\mathrm{sp}$ value can be kept fixed during the fit or it can be used as a Gaussian prior. $I_\mathrm{star}$ is a fitting parameter. The stellar radius $R_\mathrm{star}$ is taken from isochrones in every iterational steps in the following way: the effective temperature and metalicity of the star is known from spectroscopic measurements while the mean stellar density can be obtained from the scaled semi-major axis $a/R_\mathrm{star}$ which is strongly related to the transit duration \citep{seager03,winn10,csizmadia15}:
\begin{equation}
    \rho_\mathrm{star} = \frac{3\pi}{GP^2 (1+q)} \left( \frac{a}{R_\mathrm{star}}\right)^3 
\end{equation}
Then, the $R_\mathrm{star}$ value is given by the corresponding isochrones which are obtained from $\rho_\mathrm{star}$, $T_\mathrm{eff}$ and metalicity as described in \cite{csizmadia2020}.

We fit two angles: the inclination of the stellar rotational vector and its longitude of the node. These, and the planet's sky-projected position, define what is the local temperature behind the planetary disc. Then this temperature is converted via flux convolving the response function of TESS with the spectral library of Munari et al. (2005). (Such conversions are also available for CoRoT, Kepler/K2 and CHEOPS in TLCM now.) The light loss due to transits is given by the Mandel-Agol routines corrected for the gravity darkening by multiplying the limb-darkened intensity behind the planet's apparent center by the normalized  gravity darkened fluxes.

Note that longitude of node of the stellar rotational axis - denoted by $\Omega_\mathrm{star}$ - is related to angle between the projected view of the stellar rotational axis and planetary orbit's angular momentum vector via $\cos \lambda = \pm  \cos(\Omega_\mathrm{planet} - \Omega_\mathrm{star})$. Since we have set $\Omega_\mathrm{planet} = 90^\circ$ for sake of simplicity, we have $\lambda = 90^\circ - \Omega_\mathrm{star}$. (The modelling is invariant against this transformation because only the difference between the longitudes of the nodes can be measured from photometry, so one can fix one of them.\footnote{These equations are directly stemming from the unnumbered equations of Section 2.2 of \citep{csizmadia2020}.} \cite{barnes11} pointed out that photometry does not distinguish between prograde or retrograde rotation of the host stars therefore there is a degeneracy in the modelling results. According to their analysis, the following scenarios are also possible if one gets $\Omega_\mathrm{star}$ as solution: $360^\circ-\Omega_\mathrm{star}$, $180^\circ-\Omega_\mathrm{star}$,  $180^\circ+\Omega_\mathrm{star}$ as directly follows from the aforemementioned expression of $\cos \lambda$.

\subsection{Validation of the gravity darkening apporach} \label{sec:gravity_darkeing_validation}

The gravity darkening part of TLCM was tested on the first 18.2 days long data set of Kepler-13Ab. This sanity check used the SAP-FLUX data of Kepler, which were cleaned by a floating median box-car filter. We selected the data points in the $\pm 0.14 \times P$ (P is the orbital period) for this check. We had in total 11,169 data points. We fitted these data with a cosine-like baseline variation (because the reflection effect is well developed in Kepler-13A) and we set the same period, effective temperature ($T_{eff} = 8600 K$) and the contamination value what \cite{szabo20} did. We set $V\sin i_\mathrm{star} = 76.96$ km/s \citep{johnson14}. All other parameters were free.  
We compare our results to the ones of \cite{szabo20} and \cite{johnson14}:

\begin{itemize}
    \item  [$i_\mathrm{star}$:] \cite{szabo20} found - from Kepler and TESS several years long photometry - 
    the stellar rotational axis inclination to be $i_\mathrm{star} = 102.5^\circ\pm0.8^\circ$ while we have found $i_\mathrm{star} = 103.2^\circ\pm2.7^\circ$.
    
    \item [$\lambda$:] \cite{johnson14} found - from spectroscopic Rossiter-Mclaughlin measurements - the projected stellar obliquity to be $\lambda = 58.6^\circ\pm2.0^\circ$ while we have found $\lambda = 55.9^\circ\pm13.8^\circ$.
    
\end{itemize}

It is worth noting that \cite{szabo20} fixed the value of $\lambda$ for their photometric fit at the value obtained by \cite{johnson14}, we didn't. They used all available Kepler and TESS photometry, we used only 18.2 days of Kepler data for our check. These factors explain our larger error bars. The agreement between us and others are therefore reasonable and validates the TLCM-approach. 

\begin{figure*}
    \centering
    \subfloat \centering a)  {{\includegraphics[width=7cm]{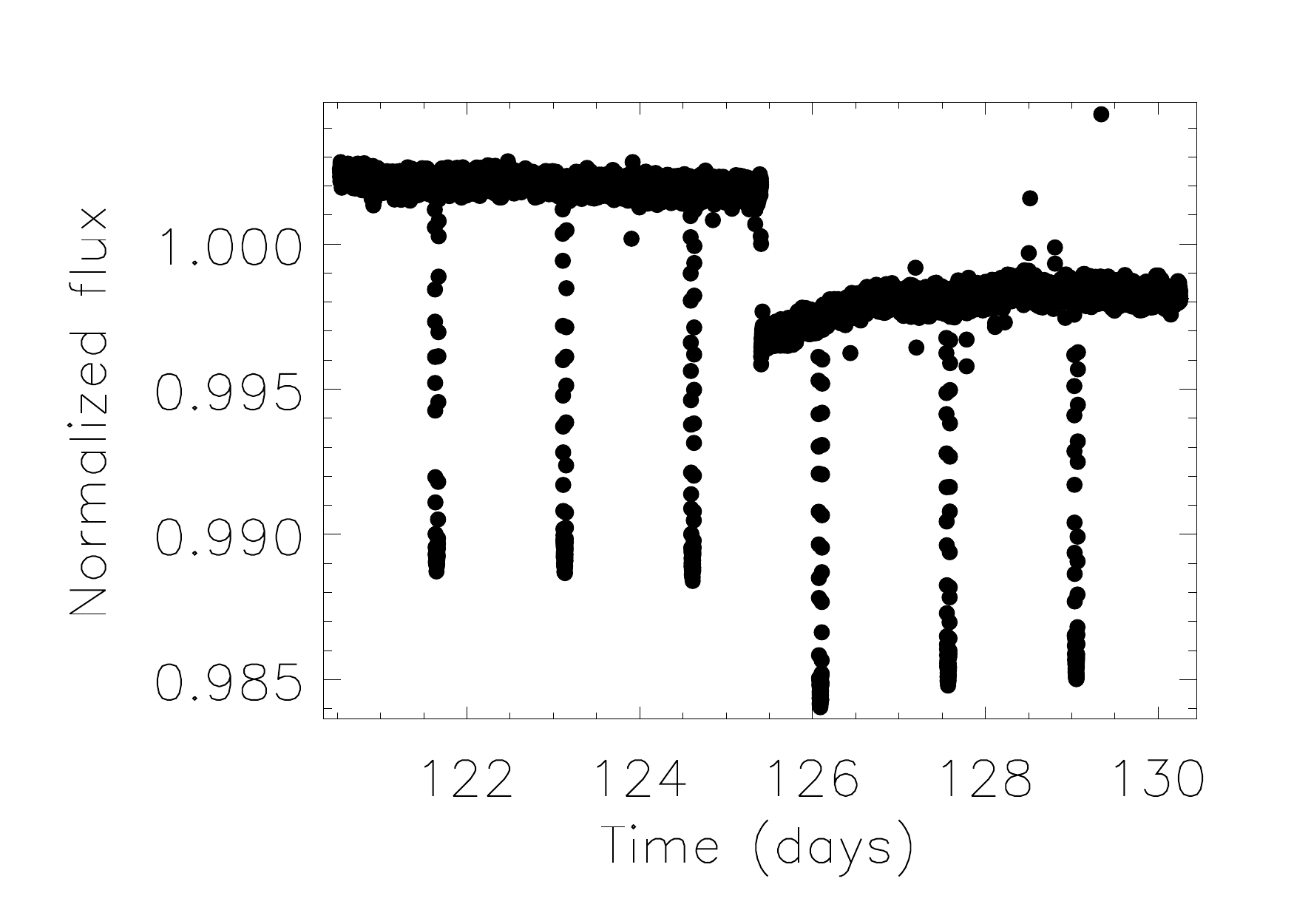} }}
    \subfloat \centering b) {{\includegraphics[width=7cm]{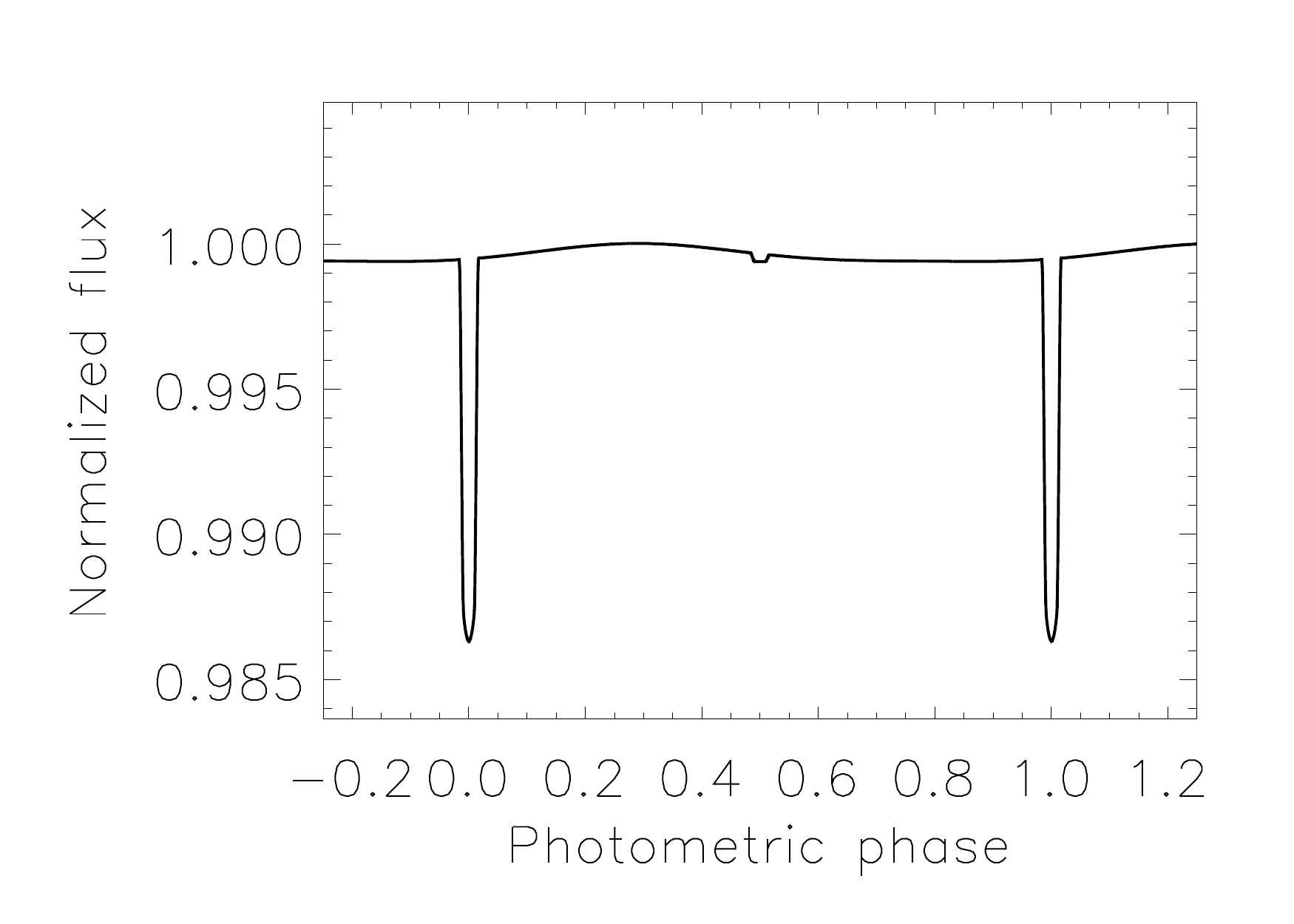}
    }} \\
        \subfloat\centering c) {{\includegraphics[width=7cm]{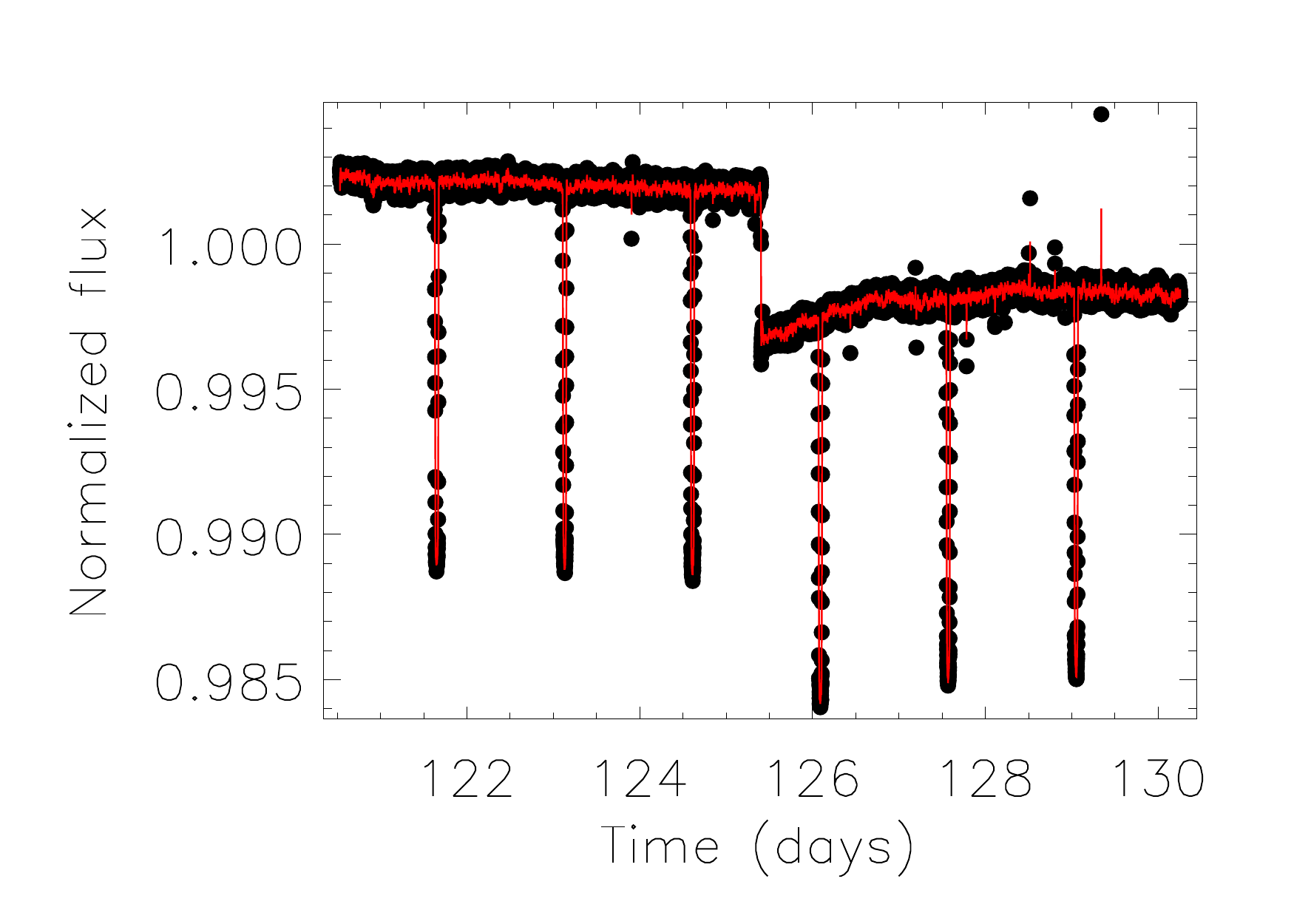}
    }}
        \subfloat \centering d) {{\includegraphics[width=7cm]{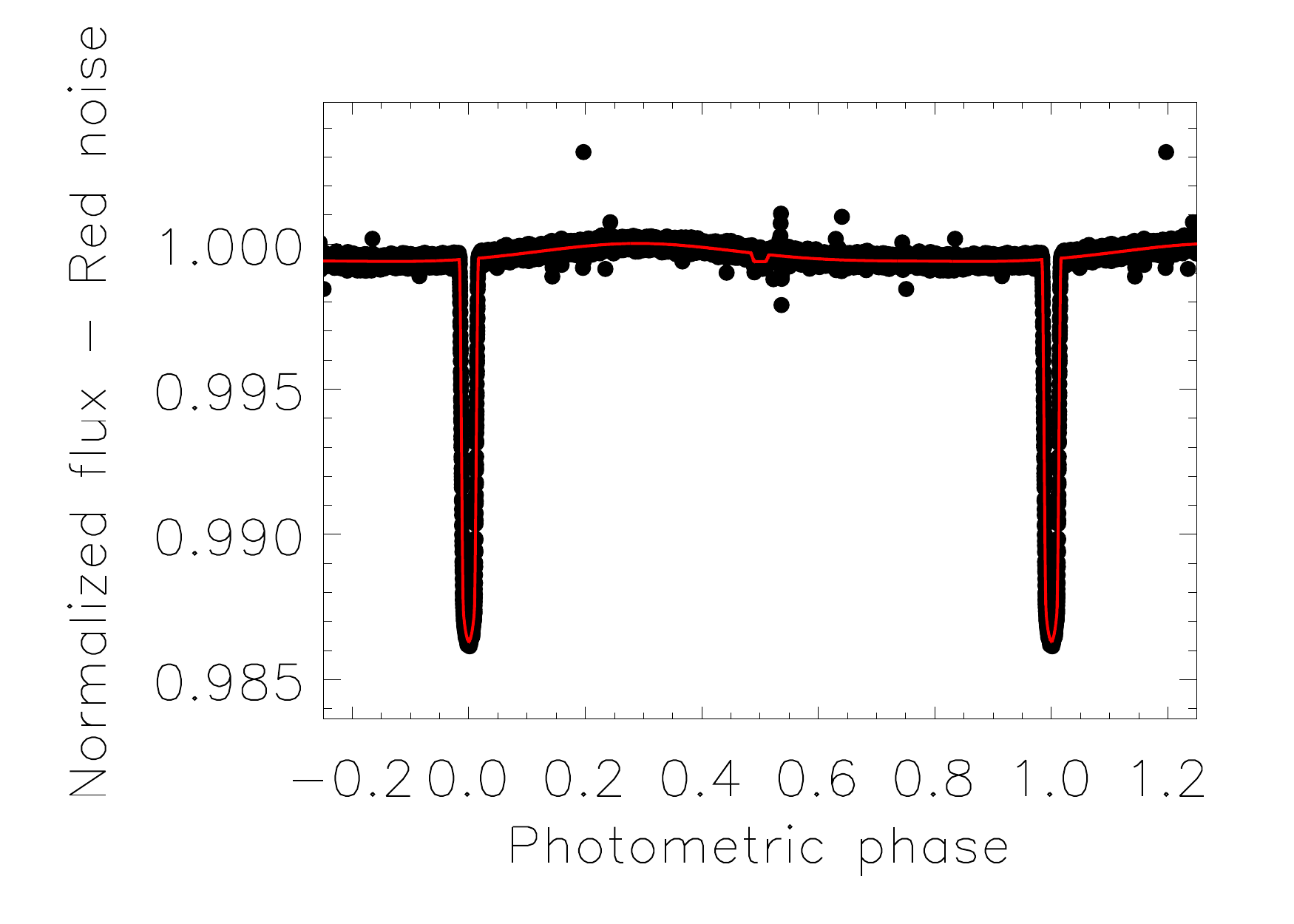}
    }}
    \caption{An example of performance of the wavelet-based light curve fit when jumps are present in the light curve. a: Raw Kepler Q1 light curve segment convolved with the injected model, used for the test (Kepler target: 001571088). b: The injected model light curve. c: Raw Kepler Q1 light curve segment (black dots) and the model$+$wavelet fit. d:
    The red noise corrected light curve (raw flux - wavelet based red noise, black dots) and the model fit (red line).
    }
    \label{fig:jump_example}
\end{figure*}

\begin{figure*}
    \centering
    \subfloat \centering a)  {{\includegraphics[width=7cm]{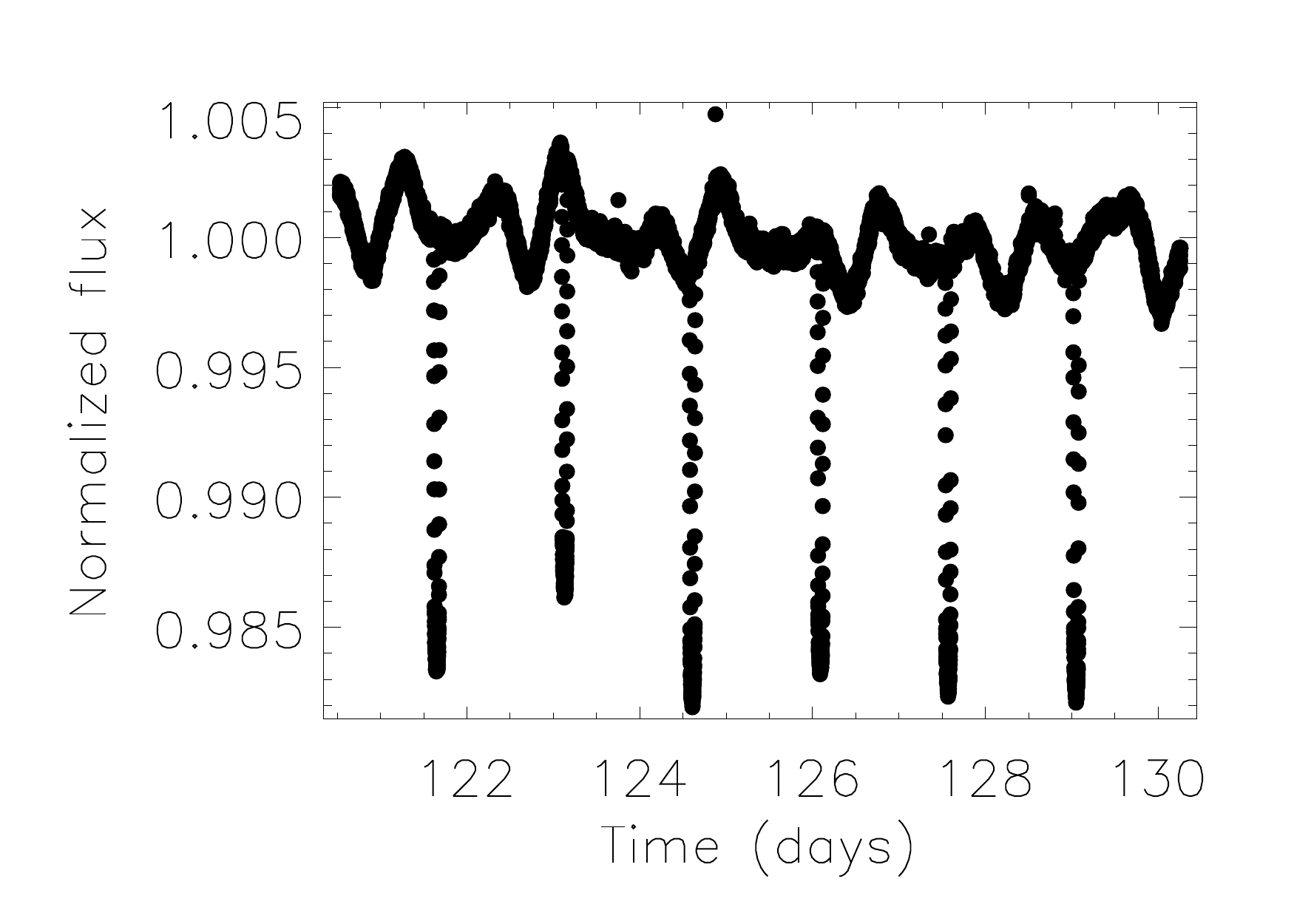} }}
    \subfloat \centering b) {{\includegraphics[width=7cm]{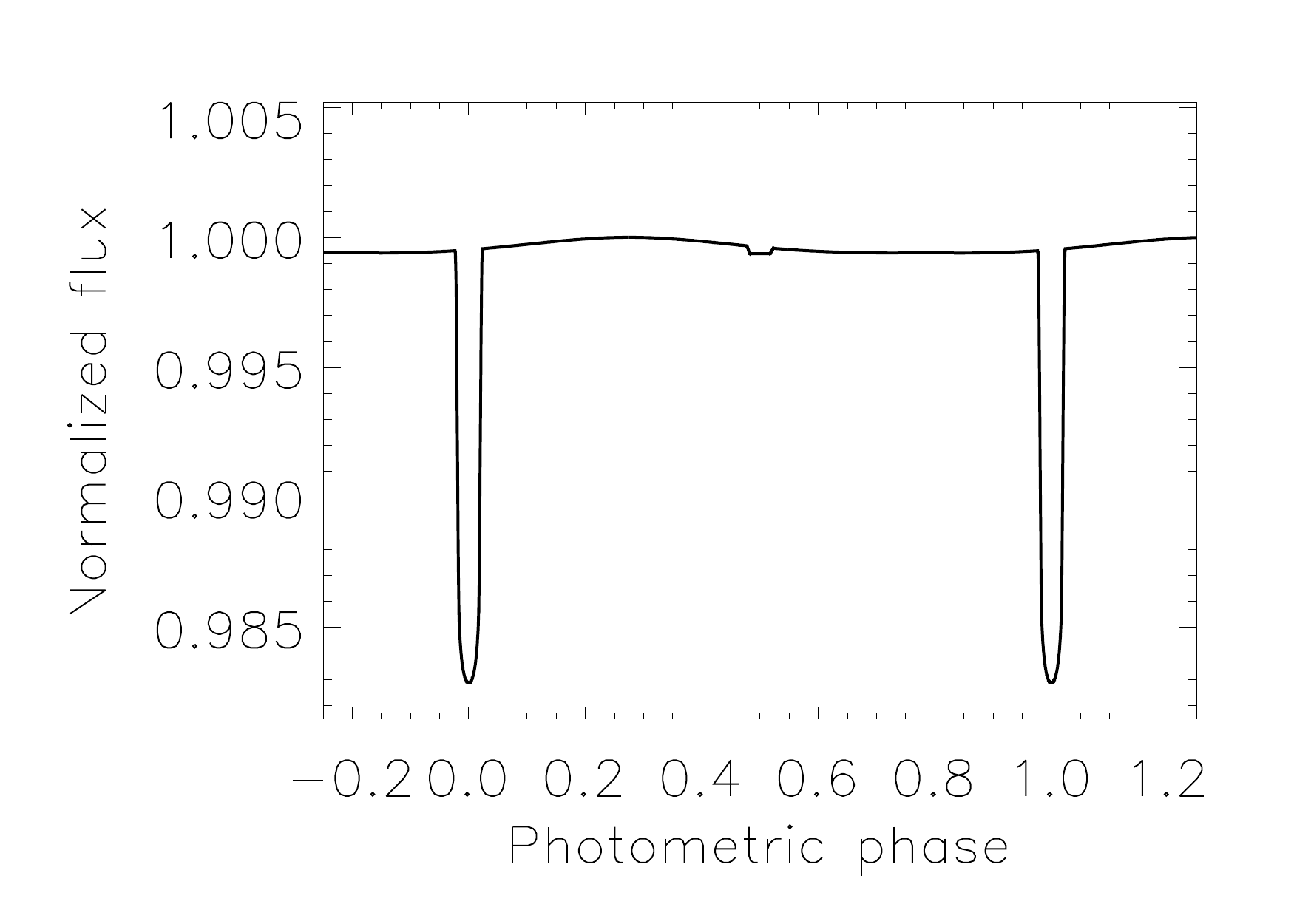}
    }} \\
        \subfloat\centering c) {{\includegraphics[width=7cm]{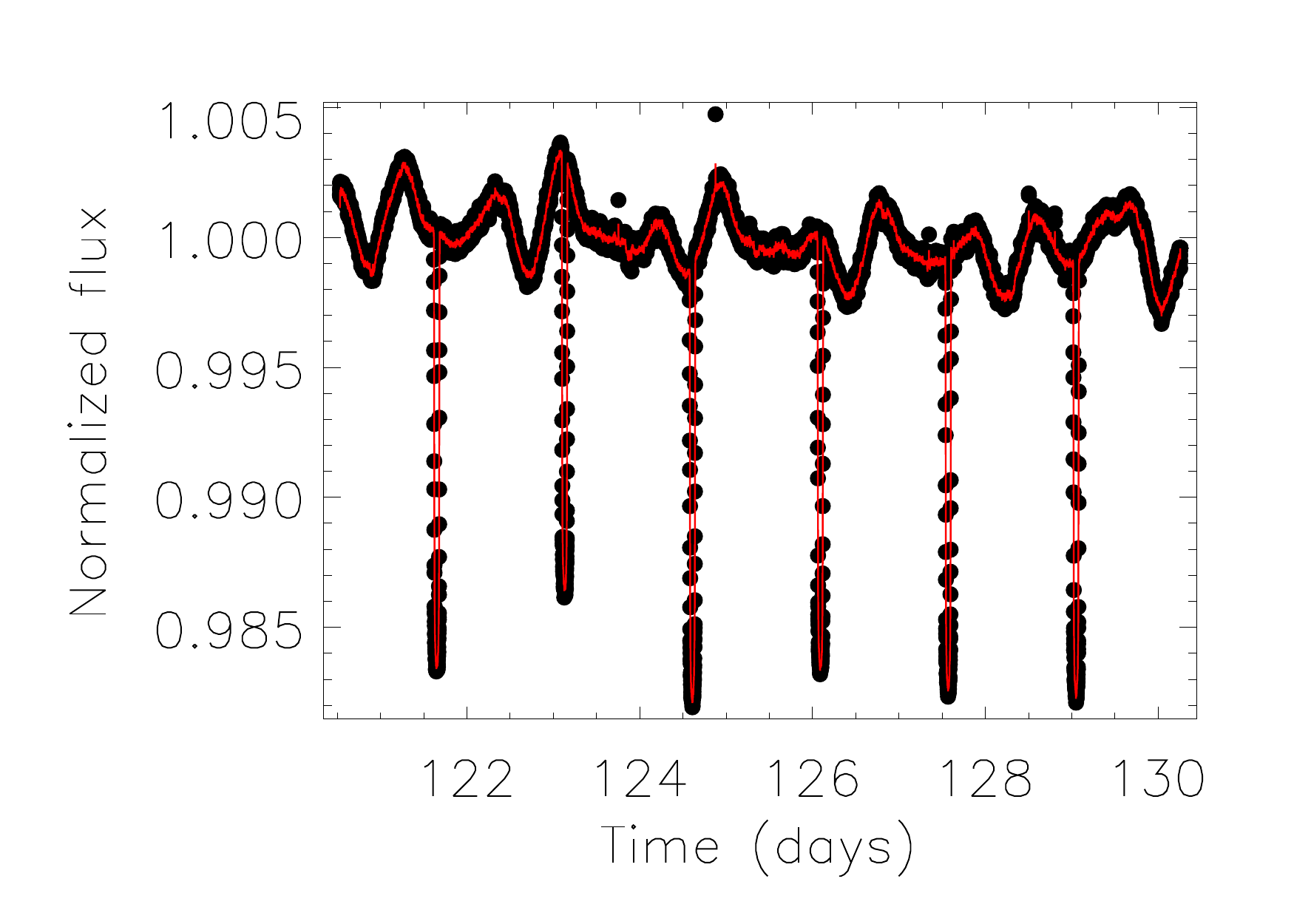}
    }}
        \subfloat \centering d) {{\includegraphics[width=7cm]{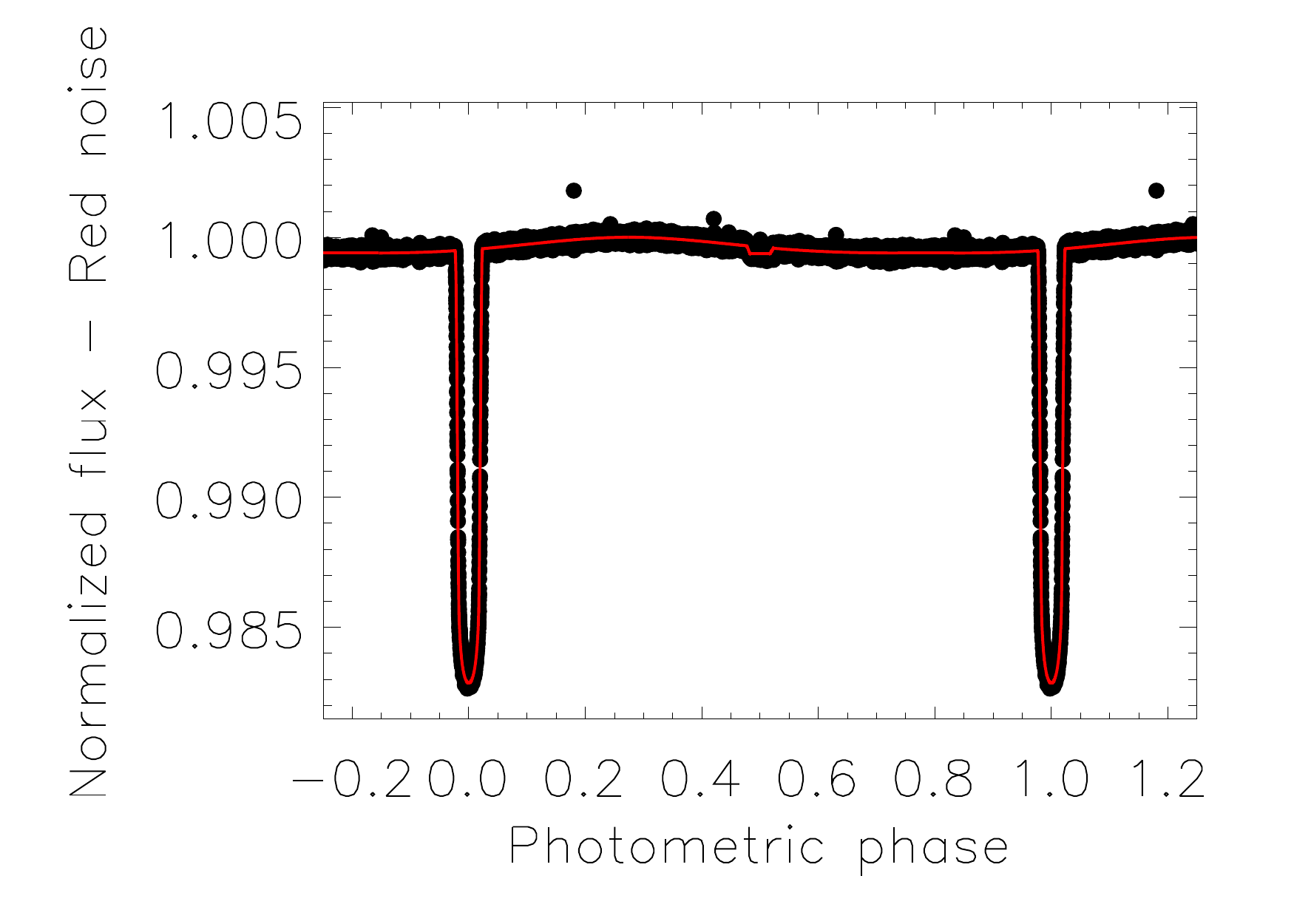}
    }}
    \caption{An example of performance of the wavelet-based light curve fit when spot-like stellar variability is present in the light curve. a: Raw Kepler Q1 light curve segment convolved with the injected model, used for the test (Kepler target: 002556755). b: The injected model light curve. c: Raw Kepler Q1 light curve segment (black dots) and the model$+$wavelet fit. d: The red noise corrected light curve (raw flux - wavelet based red noise, black dots) and the model fit (red line).
    }
    \label{fig:spot_example}
\end{figure*} 

\begin{figure*}
    \centering
    \subfloat \centering a)  {{\includegraphics[width=7cm]{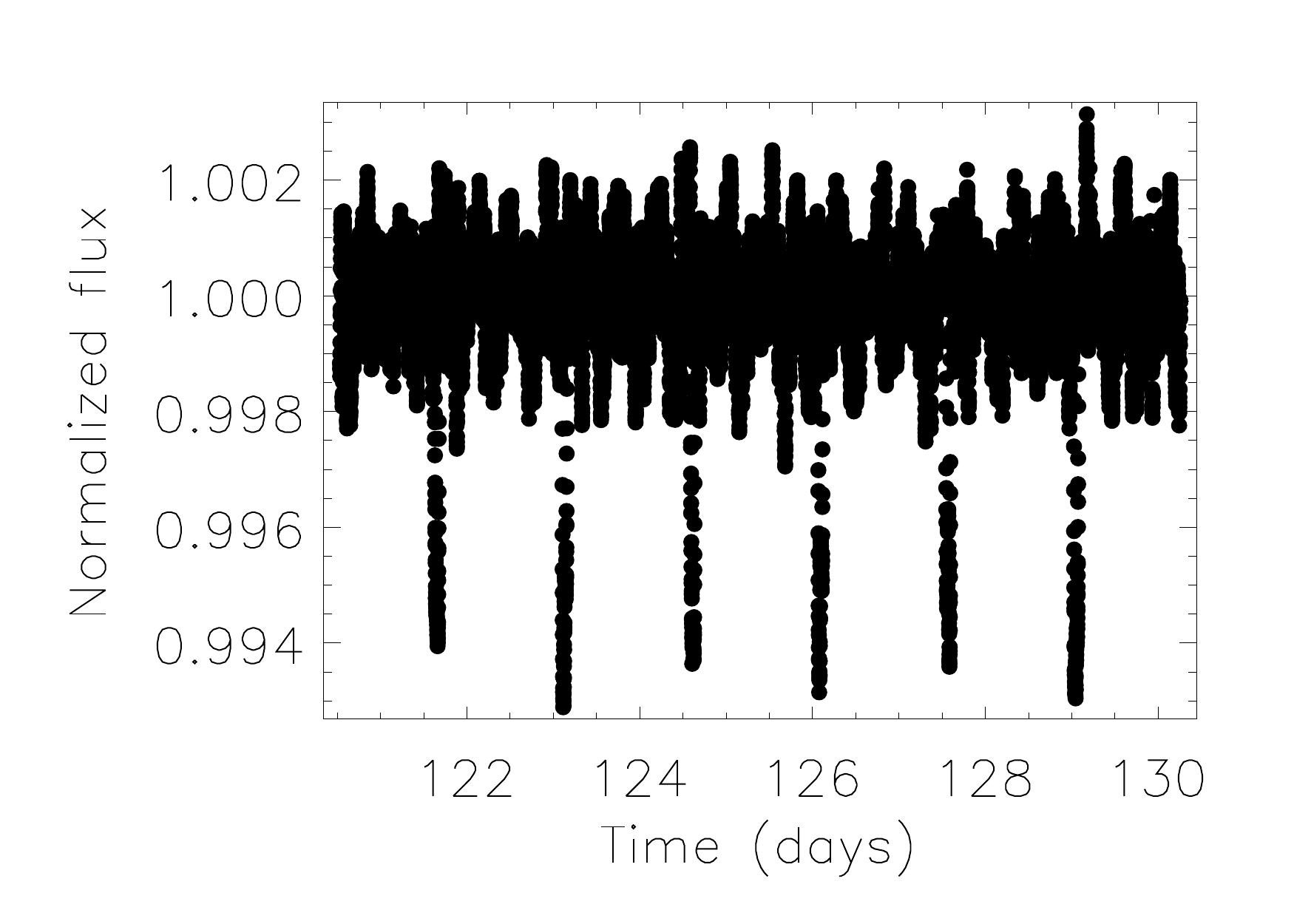} }}
    \subfloat \centering b) {{\includegraphics[width=7cm]{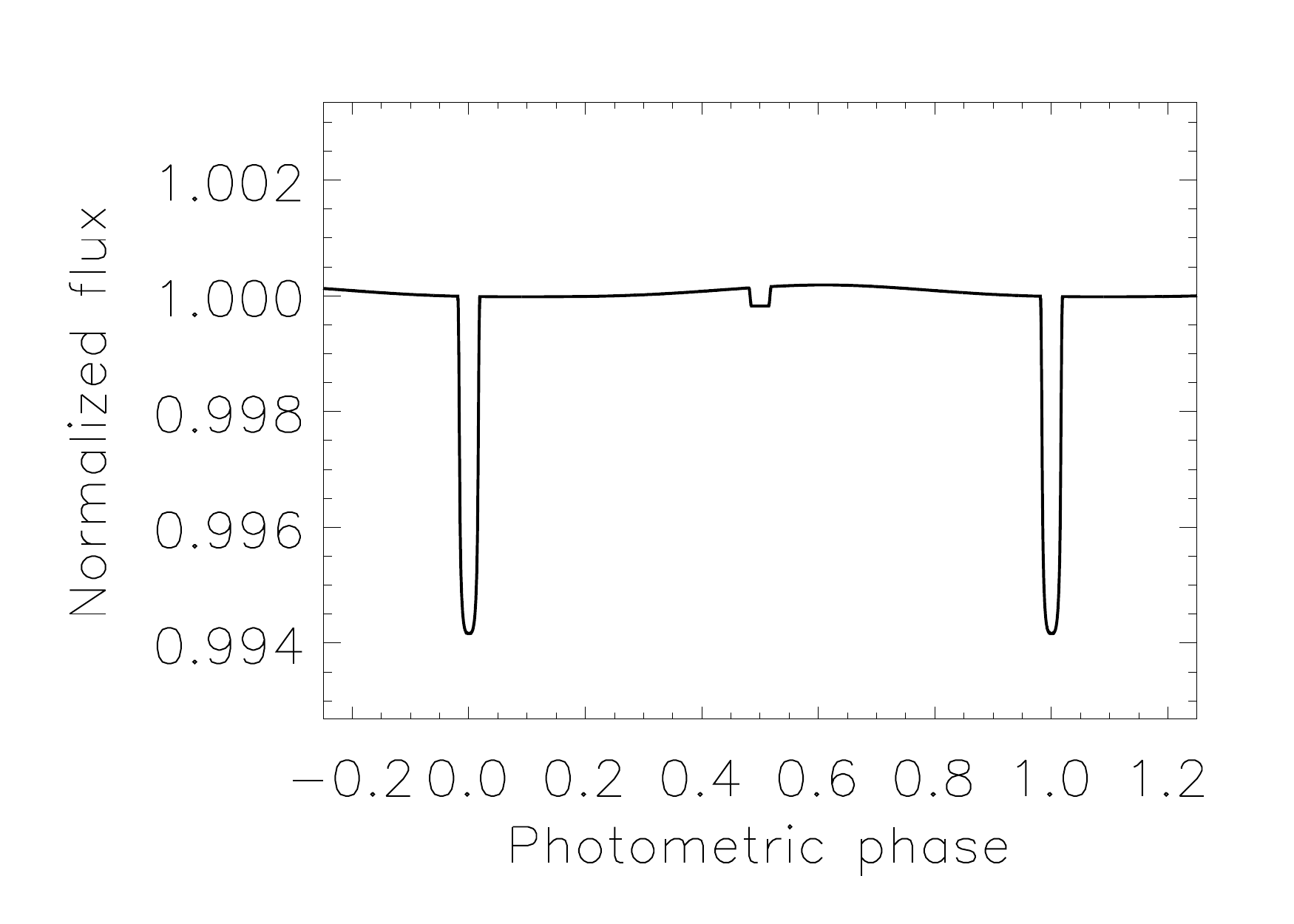}
    }} \\
        \subfloat\centering c) {{\includegraphics[width=7cm]{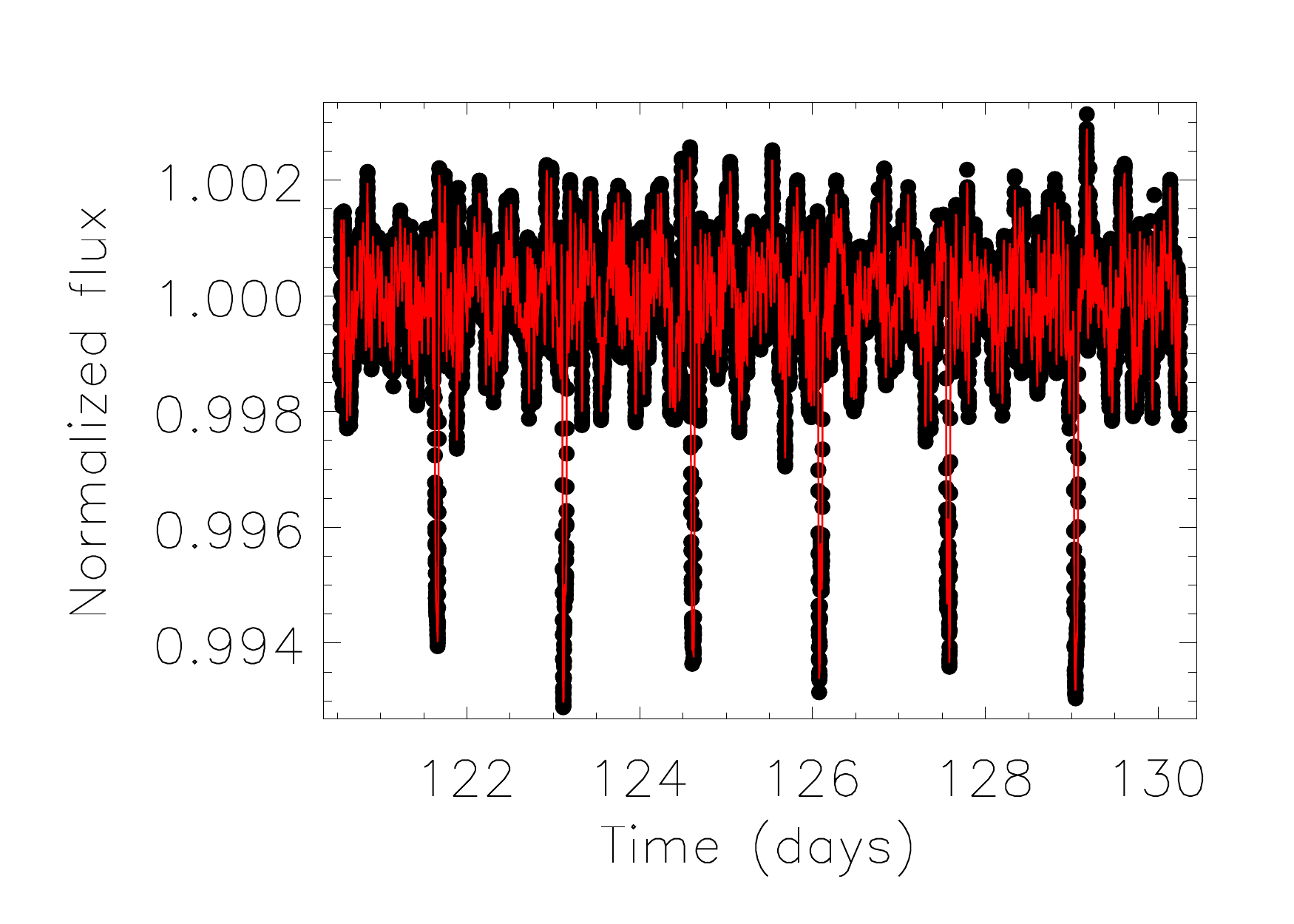}
    }}
        \subfloat \centering d) {{\includegraphics[width=7cm]{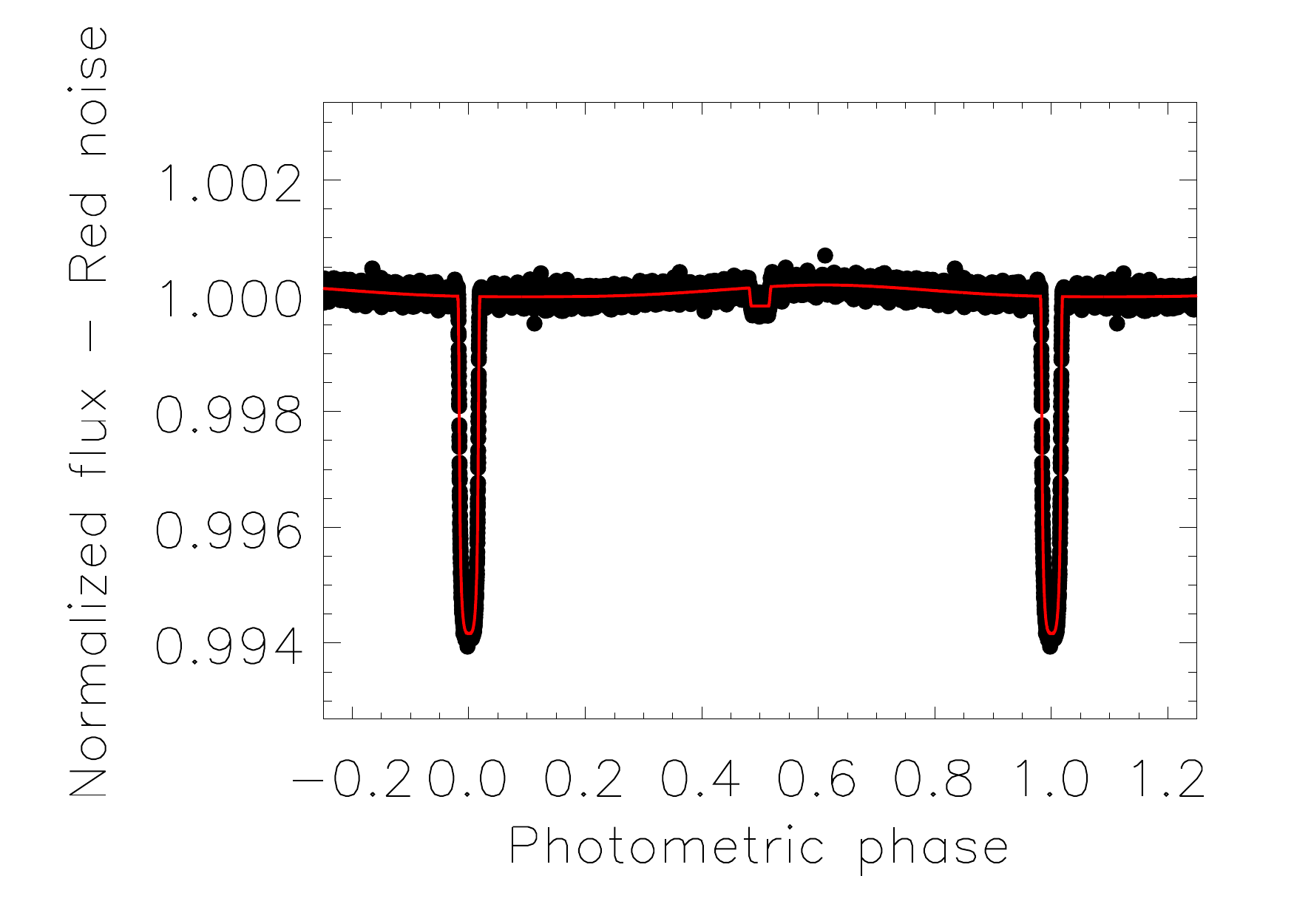}
    }}
    \caption{An example of performance of the wavelet-based light curve fit when pulsation-like stellar variability is present in the light curve. a: Raw Kepler Q1 light curve segment convolved with the injected model, used for the test (Kepler target: 004044353). b: The injected model light curve. c: Raw Kepler Q1 light curve segment (black dots) and the model$+$wavelet fit. d: The red noise corrected light curve (raw flux -- wavelet based red noise, black dots) and the model fit (red line).
    }
    \label{fig:pulsation_example}
\end{figure*}

\begin{figure*}
    \centering
    \subfloat \centering a)  {{\includegraphics[width=7cm]{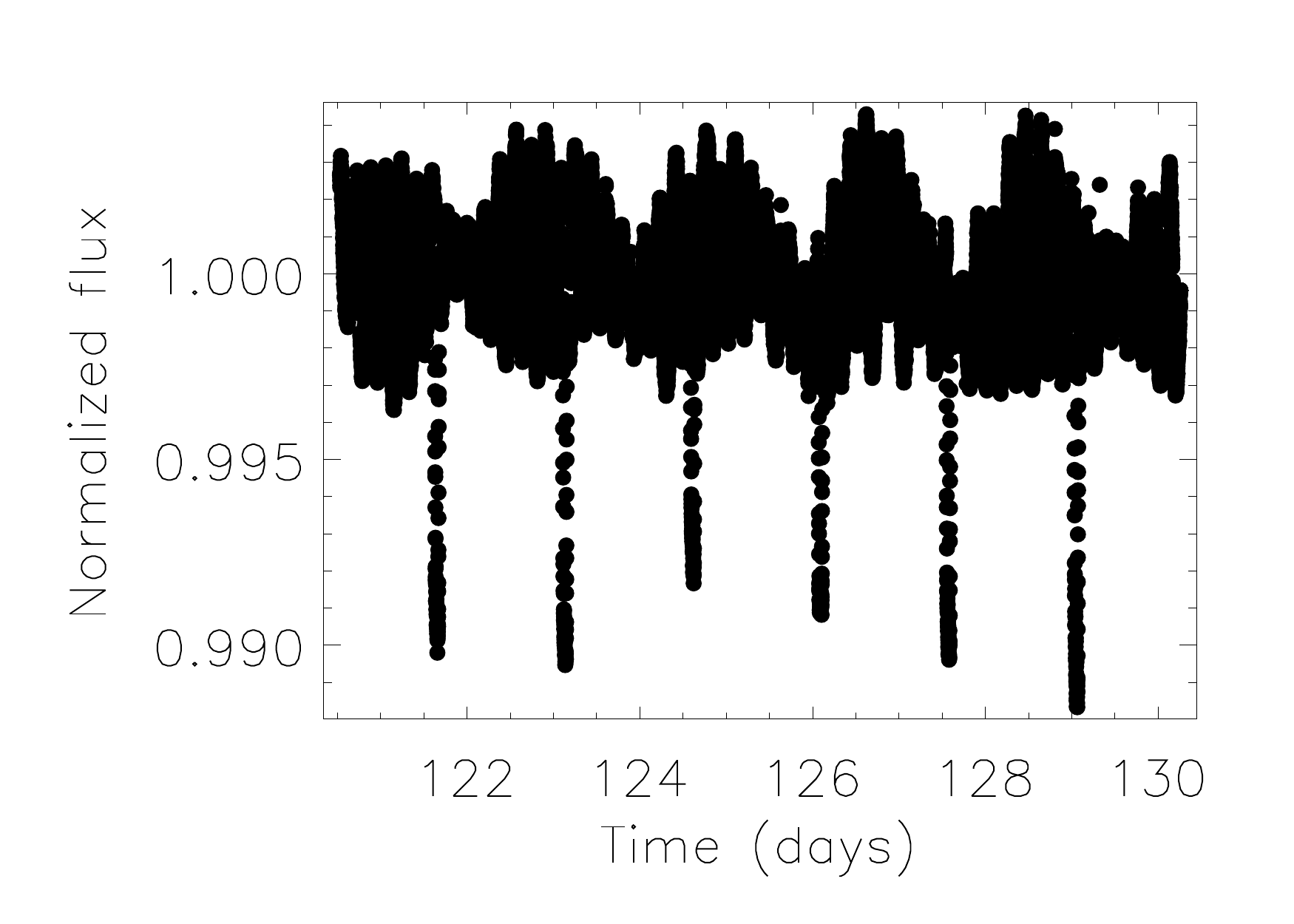} }}
    \subfloat \centering b) {{\includegraphics[width=7cm]{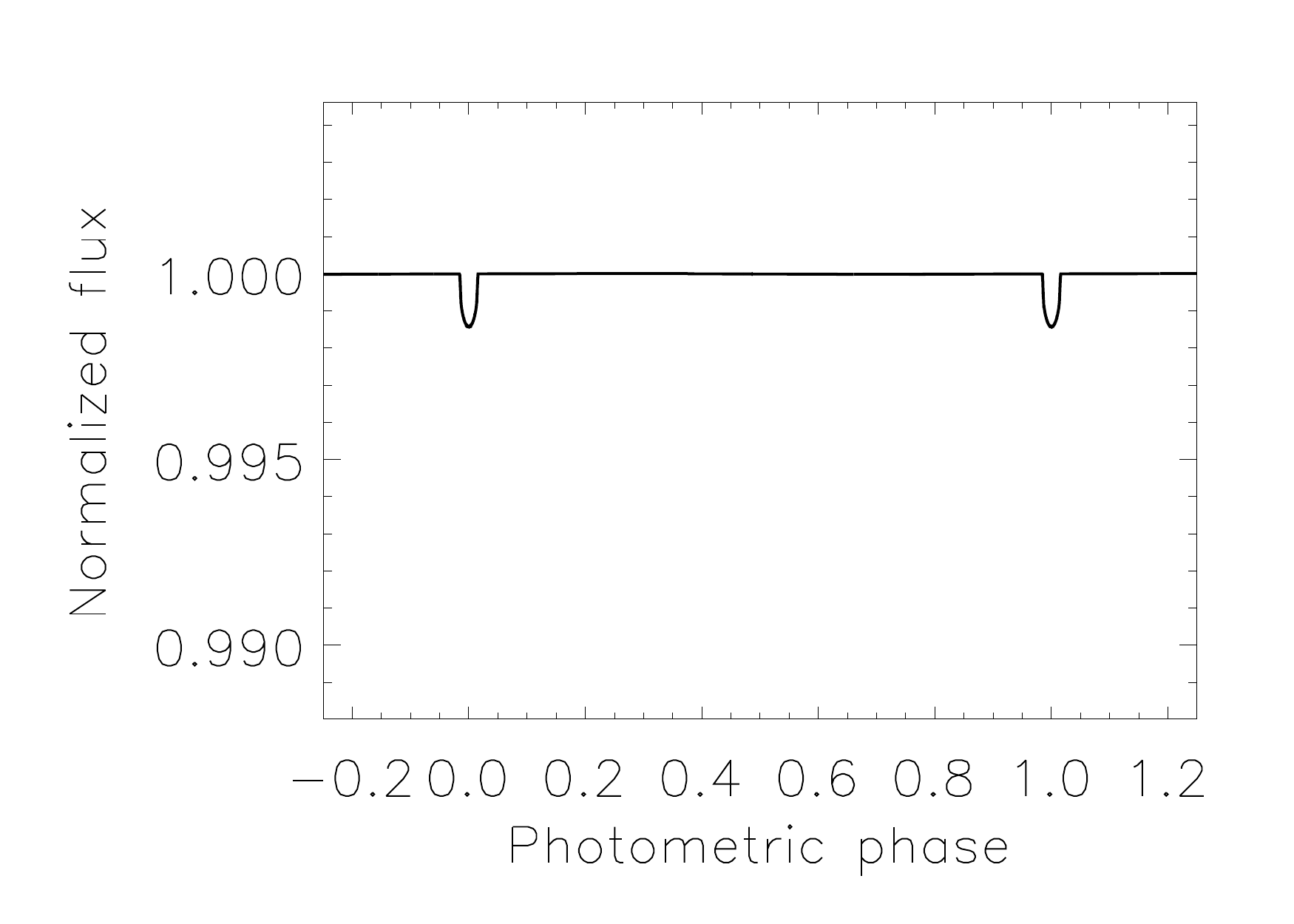}
    }} \\
        \subfloat\centering c) {{\includegraphics[width=7cm]{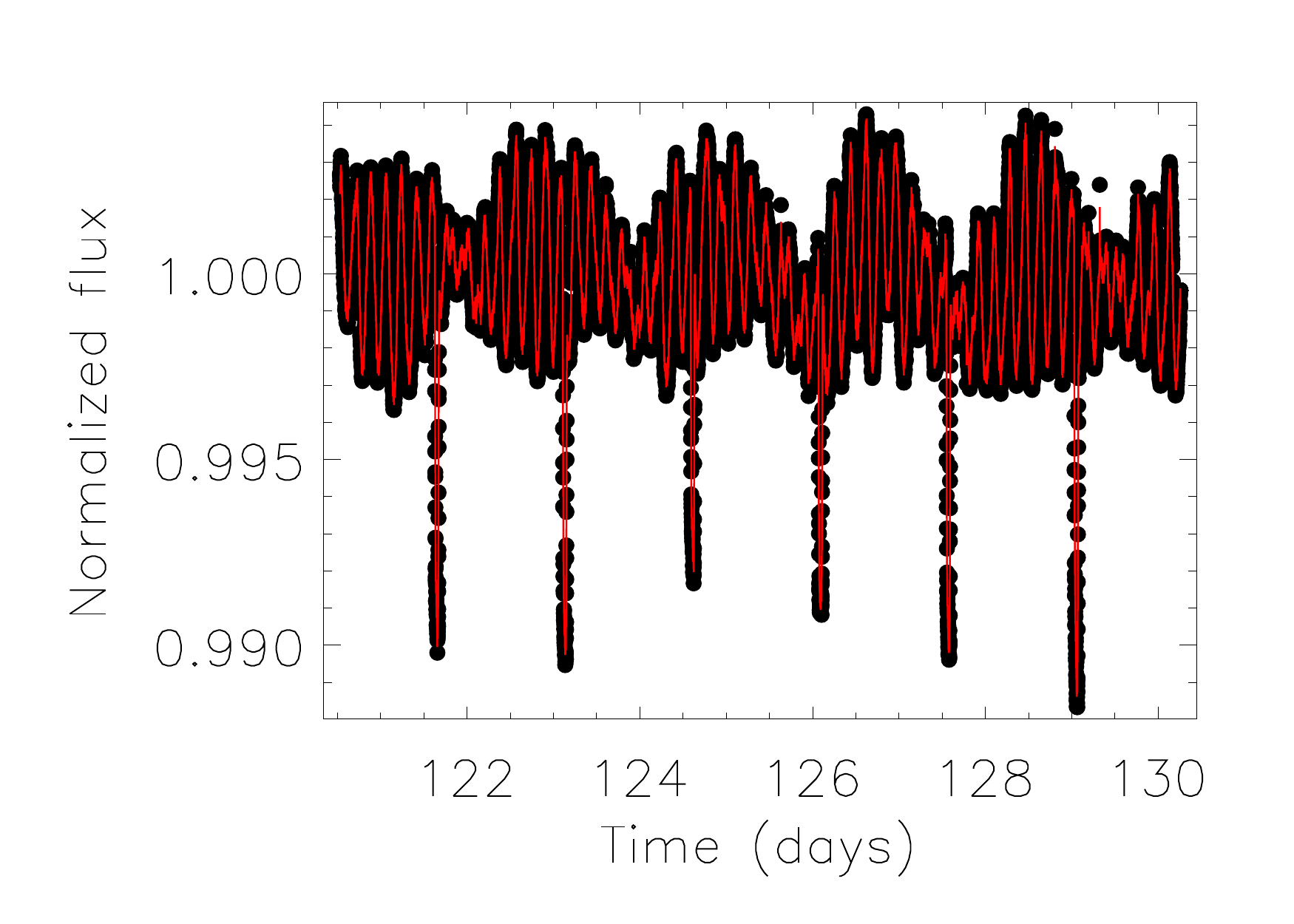}
    }}
        \subfloat \centering d) {{\includegraphics[width=7cm]{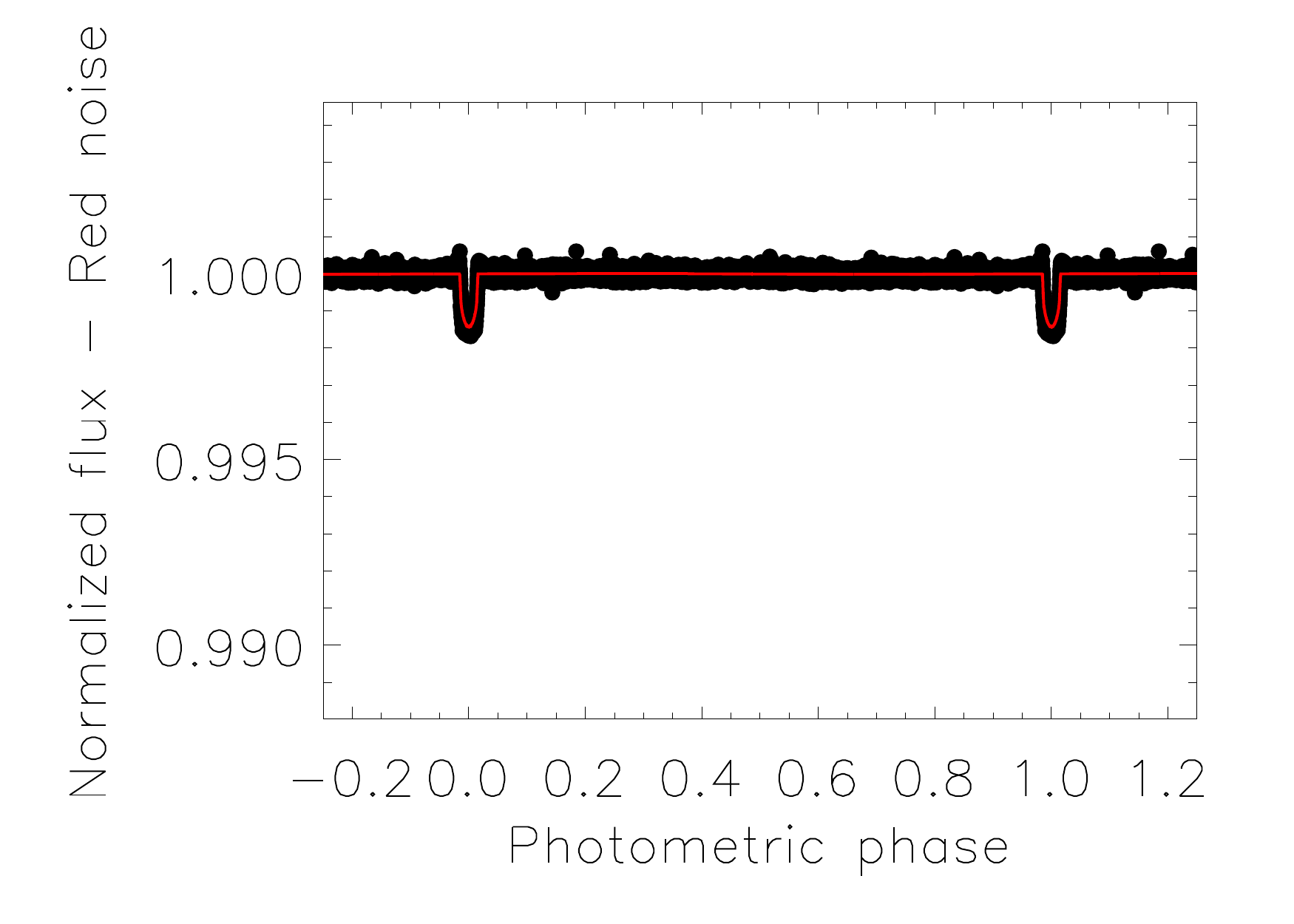}
    }}
    \caption{An example of performance of the wavelet-based light curve fit when pulsation-like stellar variability is present in the light curve. a: Raw Kepler Q1 light curve segment convolved with the injected model, used for the test (Kepler target: 010285114). b: The injected model light curve. c: Raw Kepler Q1 light curve segment (black dots) and the model$+$wavelet fit. d: The red noise corrected light curve (raw flux - wavelet based red noise, black dots) and the model fit (red line).
    }
    \label{fig:pulsation2_example}
\end{figure*}

%
\section{Wavelet-based method to remove stellar activity signals and noise-reduction} \label{sec:wavelet_method}

We used wavelets to remove any stellar activity/variability  induced signal and to reduce the noise level stemming from unknown instrumental effects. This method is also able to manage the jumps in the light curve. These jumps or data discontinuities are sudden flux increases due to a cosmic ray impact event or due to instrumental properties after rotating the satellite to re-point the solar panels or stopping observations because of data download, or telescope re-alignment etc. For instance, such data download leaks can be seen in the middle of every light curves in each sector of TESS. The mean data level shift of the same target between sectors of TESS or quarters of Kepler may due to different satellite rotation, different pointing and different contamination level and for us act as flux-jumps again.

The model of TLCM we used is based on \cite{csizmadia2020}. It is a sum of the gravity darkened transit$ + $occultation$ + $beaming$ + $reflection$ + $ellipsoidal variations$ + $wavelet based red noise model $ + $ radial velocity curve if this latter one also is available. The parameters of the different effects are fitted simultaneously.

The wavelet model is based on the work of \cite{carterwinn09}. This needs only two parameters to characterize the red (or pink) noise present in the light curve: the white noise level (root mean square, rms) $\sigma_w$ and the red-noise factor $\sigma_r$ -- this latter one is not related to the rms of the red noise component.

A difficulty in the application of the wavelets is that we do not know a priori the values of $\sigma_w$ and $\sigma_r$. When only transits and occultations are present and the out-of-transit light curve part is free of any of beaming, reflection or ellipsoidal variation, then the model gives a normalized flux = 1.0 for all out-of transit and out-of-occultation point. Then the difference between the model and the observations can be used to estimate the wavelet parameters. However, we do not have any points where we know a priori the model flux parameters if out-of-transit variations are present, except the normalization point at phase 0.25. This one point is not enough for the estimation of $\sigma_w$, $\sigma_r$.

Therefore we fit the wavelet-parameters simultaneously with the free system parameters and we apply a penalty function (prior) in the fit. The penalty function was based on the requirement that the one-sigma scatter of the fit's residuals must be equal to the average uncertainties of the photometric points. Mathematically, this meant the followings. The residual curve is defined as
\begin{equation}
    r_i = O_i - M_i
\end{equation}
where $O_i$ and $M_i$ are the observed and system model fluxes for the $i$th observations, respectively. $r_i$ is the residual of the $i$th data point. Then we calculate the loglikelihood of the noise model. To do that, we transfer the noise parameters $\sigma_w$, $\sigma_r$ and all $r_i$ values to the routines and algorithm of \cite{carterwinn09}. These routines return with the loglikelihood of the noise model (the definition of this likelihood can be found in \citealt{carterwinn09}). This loglikelihood is penaltized as
\begin{equation}
    -\log L_\mathrm{new} = -\log L +  0.5 \times N_\mathrm{data} \times \left( \frac{S(r_\mathrm{RN})}{M(\sigma_i)} - 1.0 \right)^2
\end{equation}
Here $-\log L_\mathrm{new}$ is the  minus loglikelihood to be minimized during the optimization process and used for the error estimation in the MCMC-analysis. $-\log L$ is the minus loglikelihood of the wavelet-fit to the residuals given by the algorithm of \citet{carterwinn09}. $N_\mathrm{data}$ is the number of data points. $S(r_\mathrm{RN})$ is the standard deviation of the residuals after removing the red noise component $RN_i$ which is also provided by the routines of \cite{carterwinn09}:
\begin{equation}
    r_\mathrm{RN,i} = O_i - M_i - RN_i 
\end{equation}
$\sigma_i$ is the photometric uncertainty of the $i$th observation, while $M(\sigma_i)$ is the mean of the uncertainties of all photometric individual uncertainties.

For the subsequent tests, let the free parameters be the scaled semi-major axis $a/R_{\mathrm{star}}$, the planet-to-star radius ratio $R_\mathrm{planet} / R_\mathrm{star}$, the impact parameter $b$, the sum and the difference of the linear and quadratic limb darkening coefficients $u_+ = u_a + u_b$ and $u_- = u_a - u_b$, the mass ratio $q = M_\mathrm{planet} / M_\mathrm{star}$, the surface brightness ratio $J$, the geometric-albedo of the planet $A_\mathrm{planet}$, the reflection shift parameter $\varepsilon$, period $P$, epoch $T_0$, and the wavelet-parameters $\sigma_w$ and $\sigma_r$. We assume a circular orbit and thus we fix $e=0$ for the tests. The radius of the star is assumed to be known and it is used as a prior in the way described in \cite{csizmadia2020}.

This approach was tested in the following way. We took 310, 10-day- long segments of 1-minute (short cadence, SC) light curves from the Kepler Q1 database. We convolved these light curves with simulated systems which exhibit all previously mentioned effects: transit, occultation, beaming, ellipsoidal and reflection effects. Then we modelled them with the aforementioned way with TLCM. We plotted the difference between the simulated parameters and the retrieved ones as a function of the signal-to-noise (S/N) ratio. For the ellipsoidal effect ($q$) we used the following expression of the S/N-ratio:
\begin{equation}
    S/N_{(q)} = \frac{q \left( \frac{R_\mathrm{star}}{a} \right)^3}{\sqrt{\sigma_w^2 + \sigma_r^2}} \times \sqrt{N}
        \label{eq:sn_ellipsoidal}
\end{equation}
where $N$ is the number of points in the light curve. For other  phase-curve parameters ($K$, $\varepsilon$, $A_\mathrm{planet}$) we have used
\begin{equation}
    S/N_{(A)} = \frac{A_\mathrm{planet} \left(\frac{R_\mathrm{planet}}{R_\mathrm{star}} \right)^2 \left( \frac{R_\mathrm{star}}{a} \right)^2}{\sqrt{\sigma_w^2 + \sigma_r^2}} \times \sqrt{N}
        \label{eq:sn_reflection}
\end{equation}
while for $J$ we used
\begin{equation}
    S/N_{(J)} = \frac{J \left( \frac{R_\mathrm{planet}}{R_\mathrm{star}} \right)^2}{\sqrt{\sigma_w^2 + \sigma_r^2}}  \sqrt{\frac{P R_\mathrm{star}}{\pi a t_{exp}}  \sqrt{ \left(1+\frac{R_\mathrm{planet}}{R_\mathrm{star}} \right)^2 - b^2} } \sqrt{N_\mathrm{occultations}}
        \label{eq:sn_surface_brightness}
\end{equation}
and for the transit parameters ($a/R_\mathrm{star}$, $R_\mathrm{planet}/R_\mathrm{star}$, b, limb darkening coefficients) we have used
\begin{equation}
    S/N = \frac{\left(\frac{R_\mathrm{planet}} {R_\mathrm{star}} \right)^2 \sqrt{\frac{P R_\mathrm{star}}{\pi a t_{exp}} \sqrt{(1+R_\mathrm{planet}/R_\mathrm{star})^2 - b^2} } }{\sqrt{\sigma_w^2 + \sigma_r^2}} \sqrt{N_\mathrm{transit}}
    \label{eq:sn_transit}
\end{equation}
where we took into account that transit and occultation duration, and hence number of in-transit (in-occultation)  points are different at different impact parameters. 

In Figures~\ref{fig:jump_example}-\ref{fig:pulsation2_example} we show some examples of the test: the injected light curve, the convolved light curve which contains all the red-noise effect induced in the Kepler-light curves, the comparison of the 'synthetic observations' and the modelling, and finally the red noise corrected light curve and the fits. 

\begin{figure}
   \centering
   \includegraphics[width=9cm]{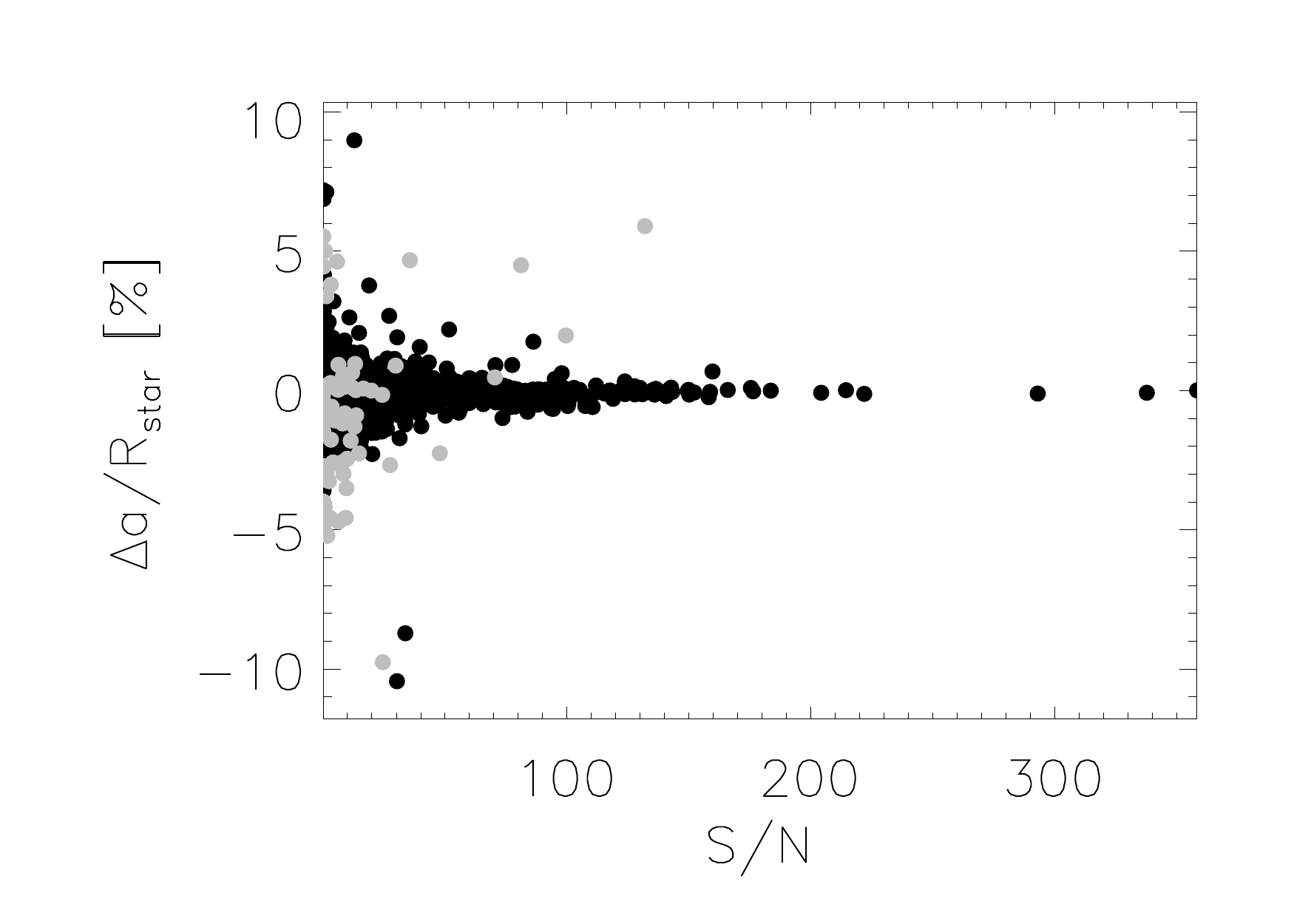}
      \caption{The result of the light curve test. The ordinate is the  S/N-ratio defined by Eq.~(\ref{eq:sn_transit}). The y-axis is the  difference between the simulated and the retrieved scaled semi-major axis in percentage. Black dots denote the solutions where the stellar radius value was obtained to be with 3\% accuracy relative to the injected stellar radius, while gray points represent the cases where we had obtained them with 3-6\% accuray. \label{fig:prove_aRstar}
      }
   \end{figure}

\begin{figure}
   \centering
   \includegraphics[width=9cm]{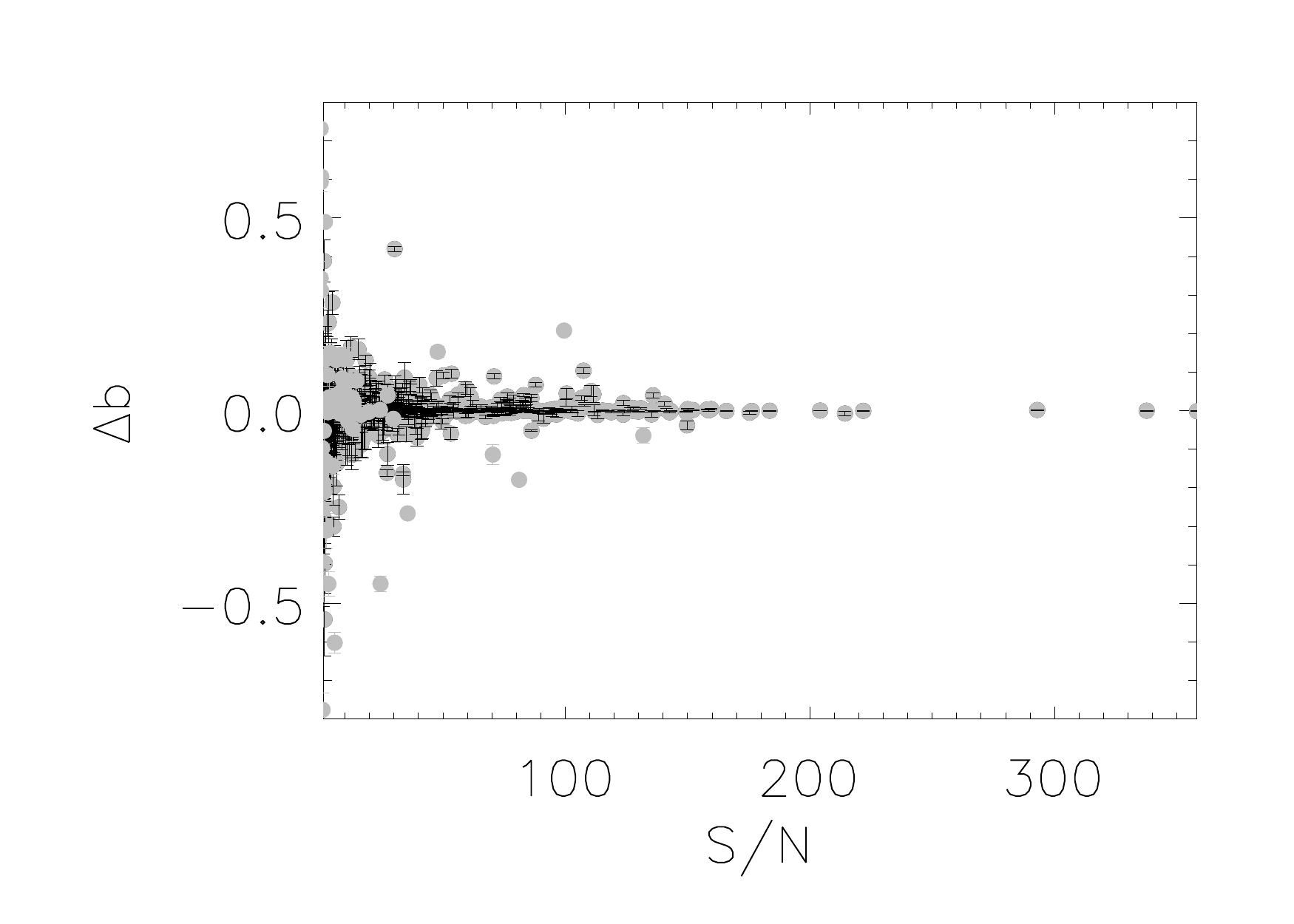}
      \caption{The result of the light curve test. The ordinate is the  S/N-ratio defined by Eq.~(\ref{eq:sn_transit}). The y-axis is the  difference between the simulated and the retrieved impact parameter. We also plotted the $1\sigma$ error bar of the impact parameter for this figure. See the meaning of black/gray points at Figure~\ref{fig:prove_aRstar}. \label{fig:prove_b}
      }
   \end{figure}

\begin{figure}
   \centering
   \includegraphics[width=9cm]{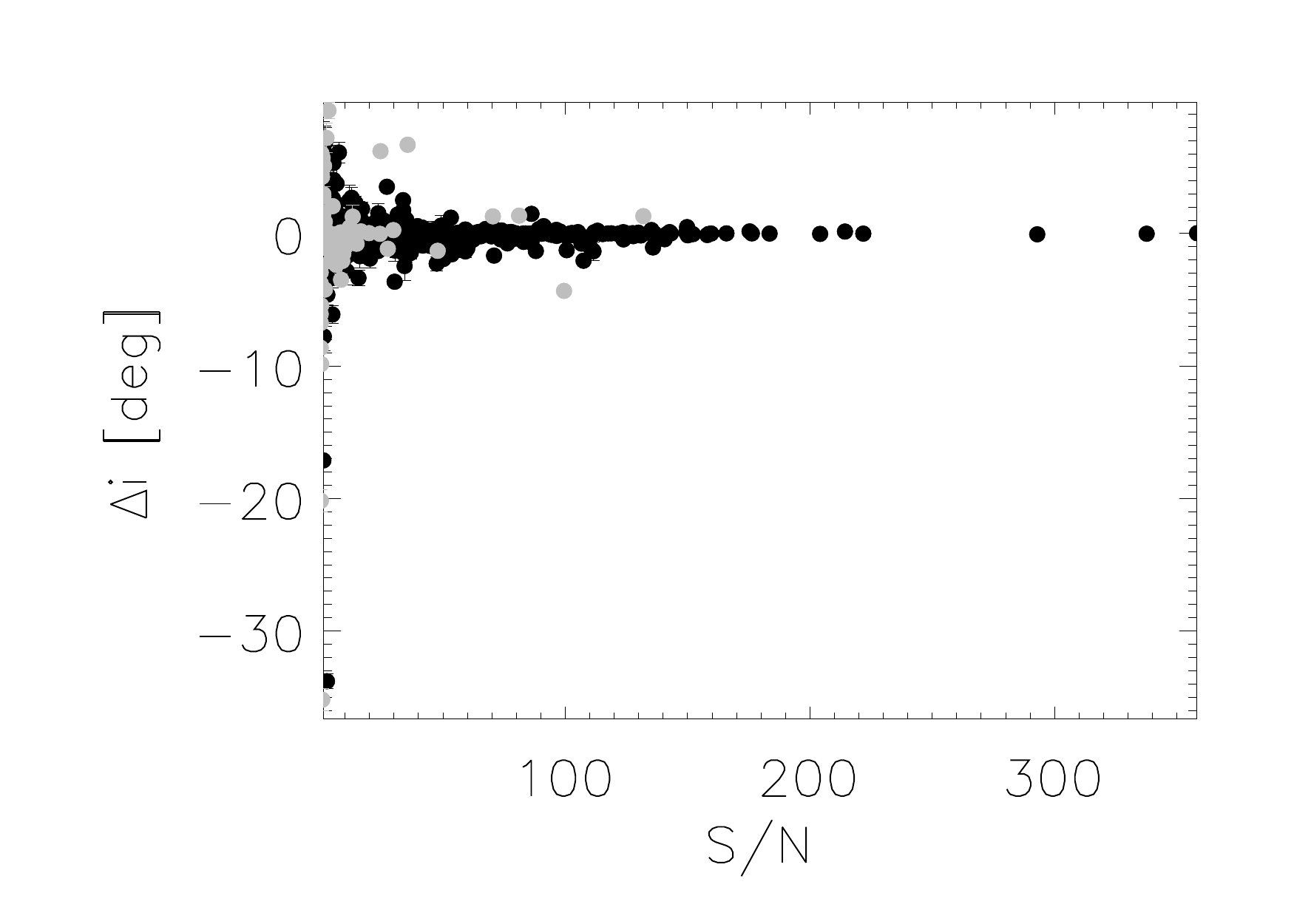}
      \caption{The result of the light curve test. The ordinate is the  S/N-ratio defined by Eq.~(\ref{eq:sn_transit}). The y-axis is the  difference between the simulated and the retrieved inclination values. The vertical lines are the $1\sigma$ error bars. See the meaning of black/gray points at Figure~\ref{fig:prove_aRstar}. \label{fig:prove_inclination}
      }
   \end{figure}

\begin{figure}
   \centering
   \includegraphics[width=9cm]{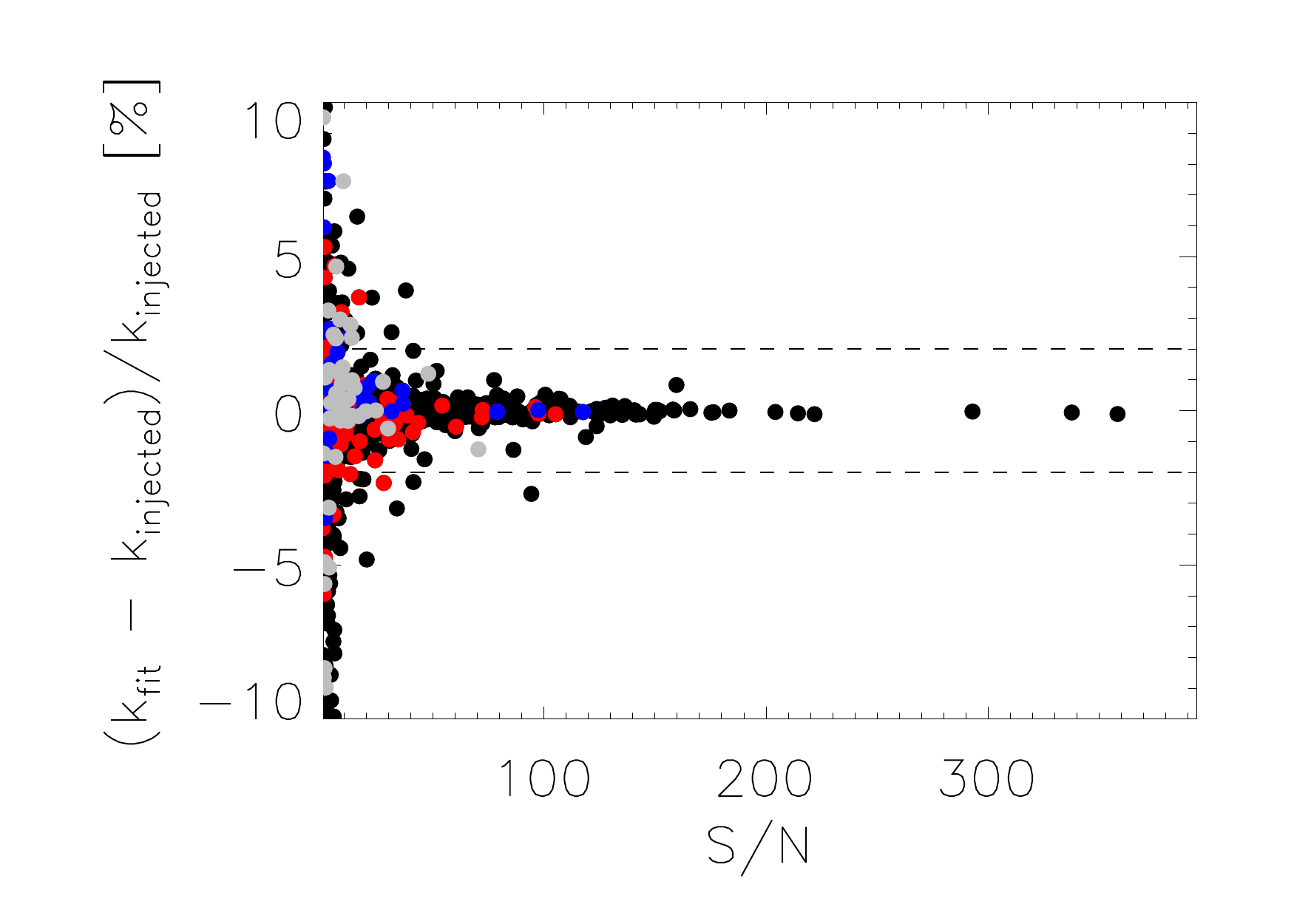}
      \caption{The result of the light curve test. The ordinate is the  S/N-ratio defined by Eq.~(\ref{eq:sn_transit}). The y-axis is the  difference between the simulated and the retrieved planet-to-star radius ratio values ($k=R_\mathrm{planet} / R_\mathrm{star}$). The red and blue points are the small superearths and earths with $0.015<k<0.03$ (red) and $0.005<k<0.015$ (blue). Note that these radius ratios correspond to Super/Earth-sized (red) and Neptune-sized (blue) planets around a solar-sized star. (The horizontal dashed lines denote $\pm2$\% relative errors in the radius ratio. For bigger companions, black dots denote the solutions where the stellar radius value was obtained to be with 3\% accuracy relative to the injected stellar radius, while gray points represent the cases where we had obtained them with 3-6\% accuray. \label{fig:prove_radiusratio}
      }
   \end{figure}

\begin{figure}
   \centering
   \includegraphics[width=9cm]{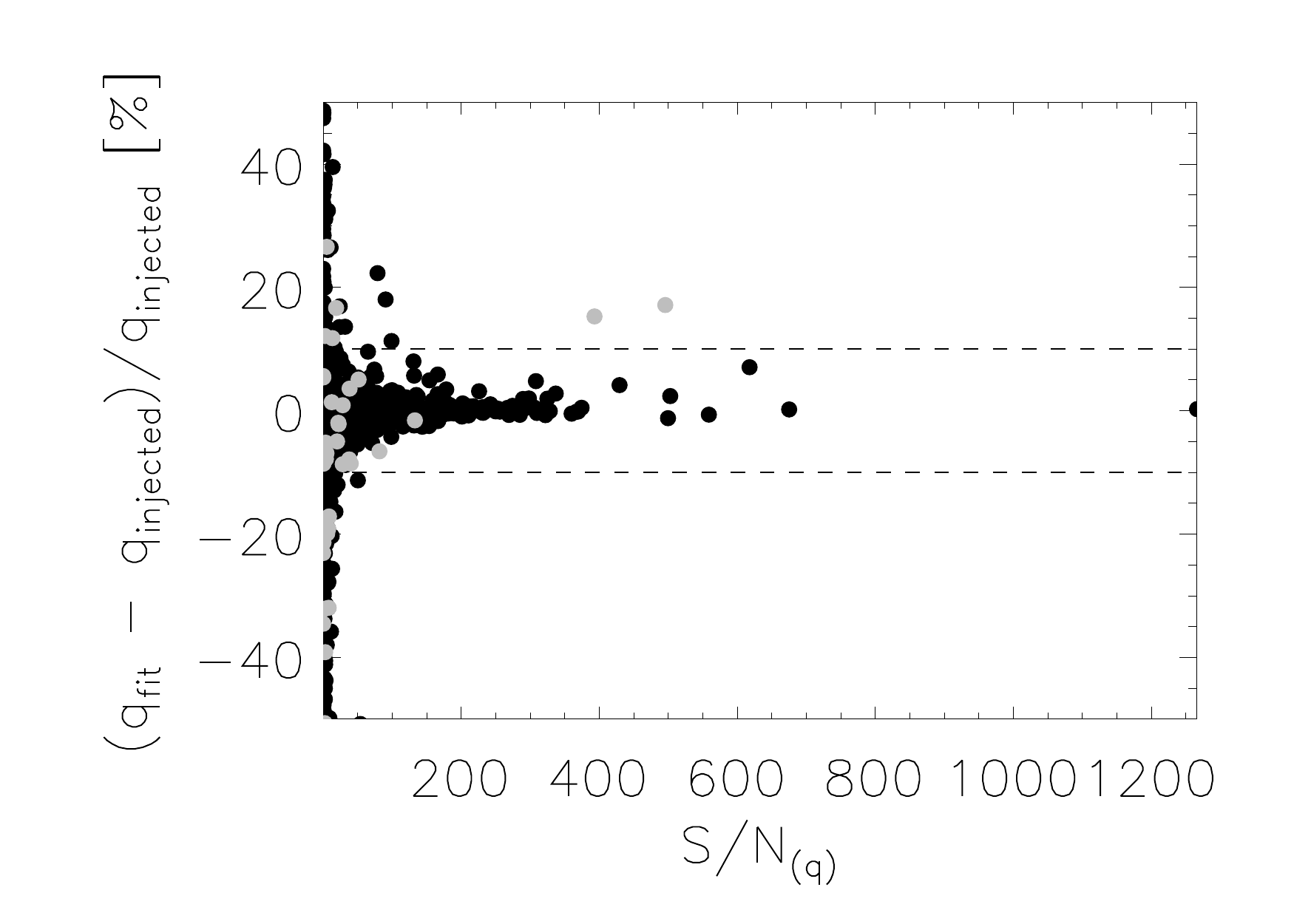}
      \caption{The result of the light curve test. The ordinate is the  S/N-ratio defined by Eq.~(\ref{eq:sn_ellipsoidal}). The abcissa is the  difference between the simulated and the retrieved planet-to-star mass ratio values. The horizontal dashed lines denote $\pm10$\% relative errors in the radius ratio.  \label{fig:prove_massratio}
      }
   \end{figure}

\begin{figure}
   \centering
   \includegraphics[width=9cm]{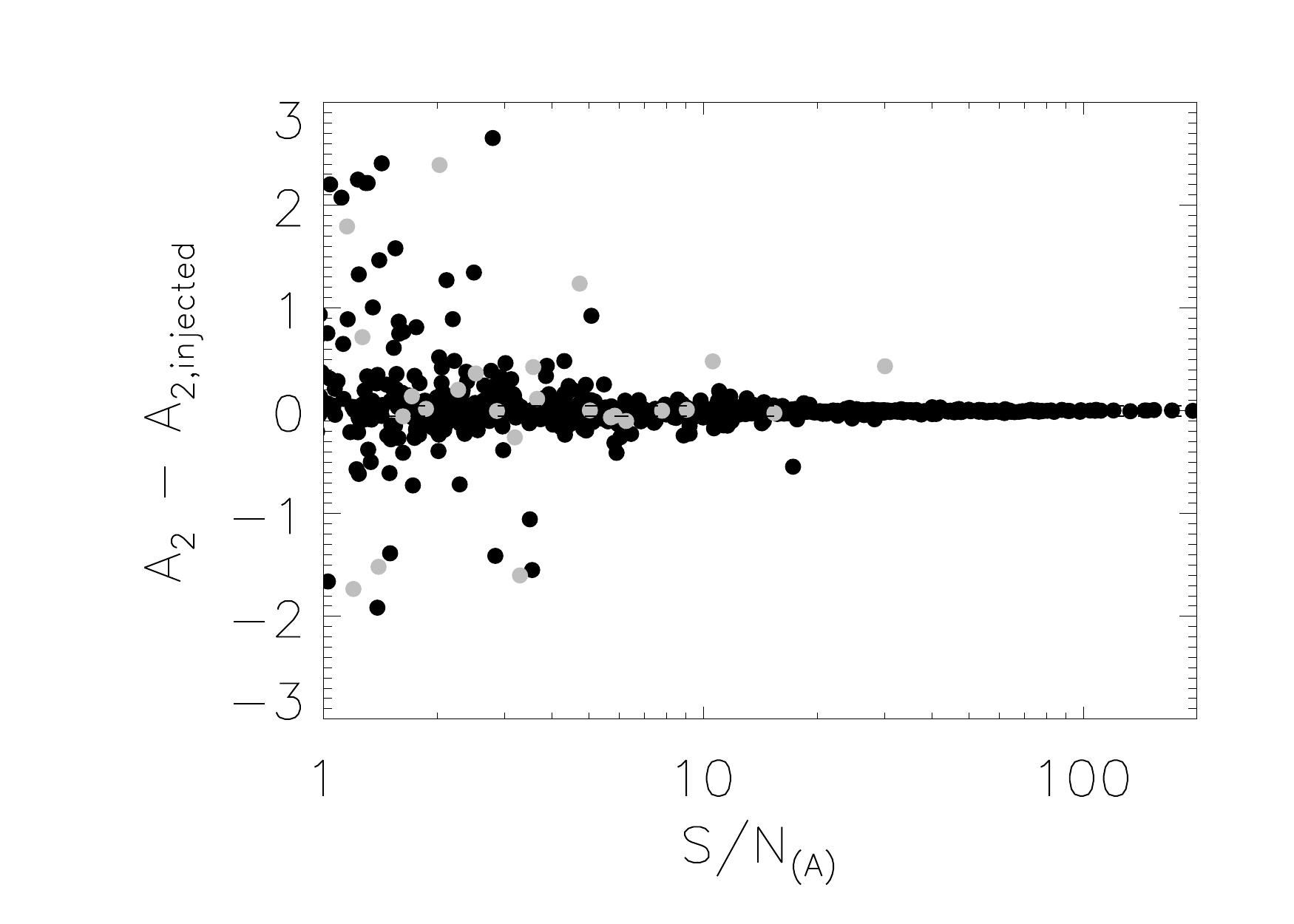}
      \caption{The result of the light curve test. The ordinate is the  S/N-ratio defined by Eq.~(\ref{eq:sn_reflection}). The y-axis is the  difference between the simulated and the retrieved planetary albedo values. The horizontal dashed lines denote $\pm0.05$ absolute errors in albedo-determination. Note that in this Figure the x-axis has a logarithmic scale for better visibility. See the meaning of black/gray points at Figure~\ref{fig:prove_aRstar}. \label{fig:prove_albedo2}
      }
   \end{figure}

\begin{figure}
   \centering
   \includegraphics[width=9cm]{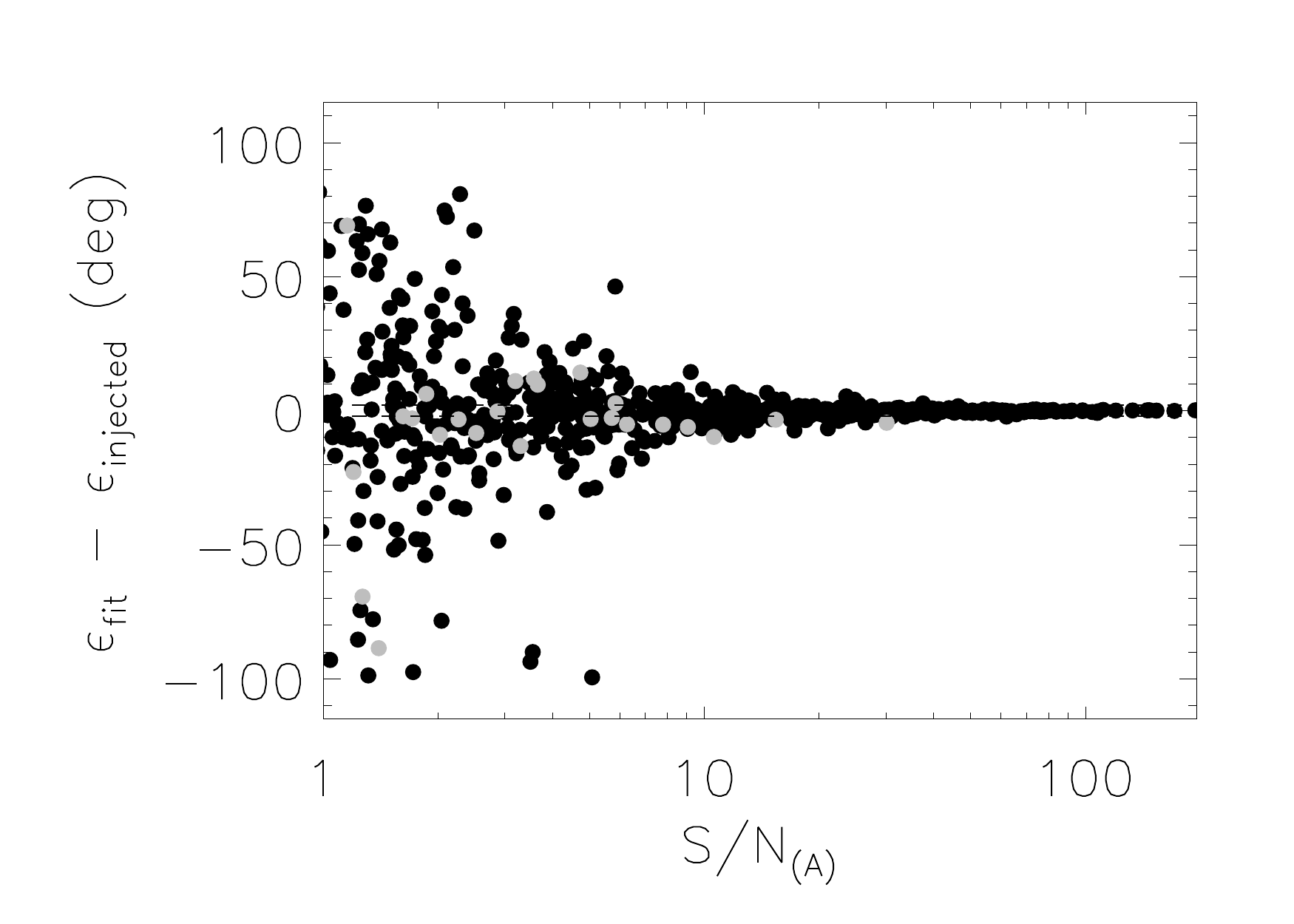}
      \caption{The result of the light curve test. The ordinate is the  S/N-ratio defined by Eq.~(\ref{eq:sn_reflection}). The abcissa is the  difference between the simulated and the retrieved reflection shift values. The horizontal dashed lines denote $\pm2^\circ$ relative errors in the radius ratio. Note that in this Figure the x-axis has a logarithmic scale for better visibility. See the meaning of black/gray points at Figure~\ref{fig:prove_aRstar}. \label{fig:prove_reflectionshift}
      }
   \end{figure}

\begin{figure}
   \centering
   \includegraphics[width=9cm]{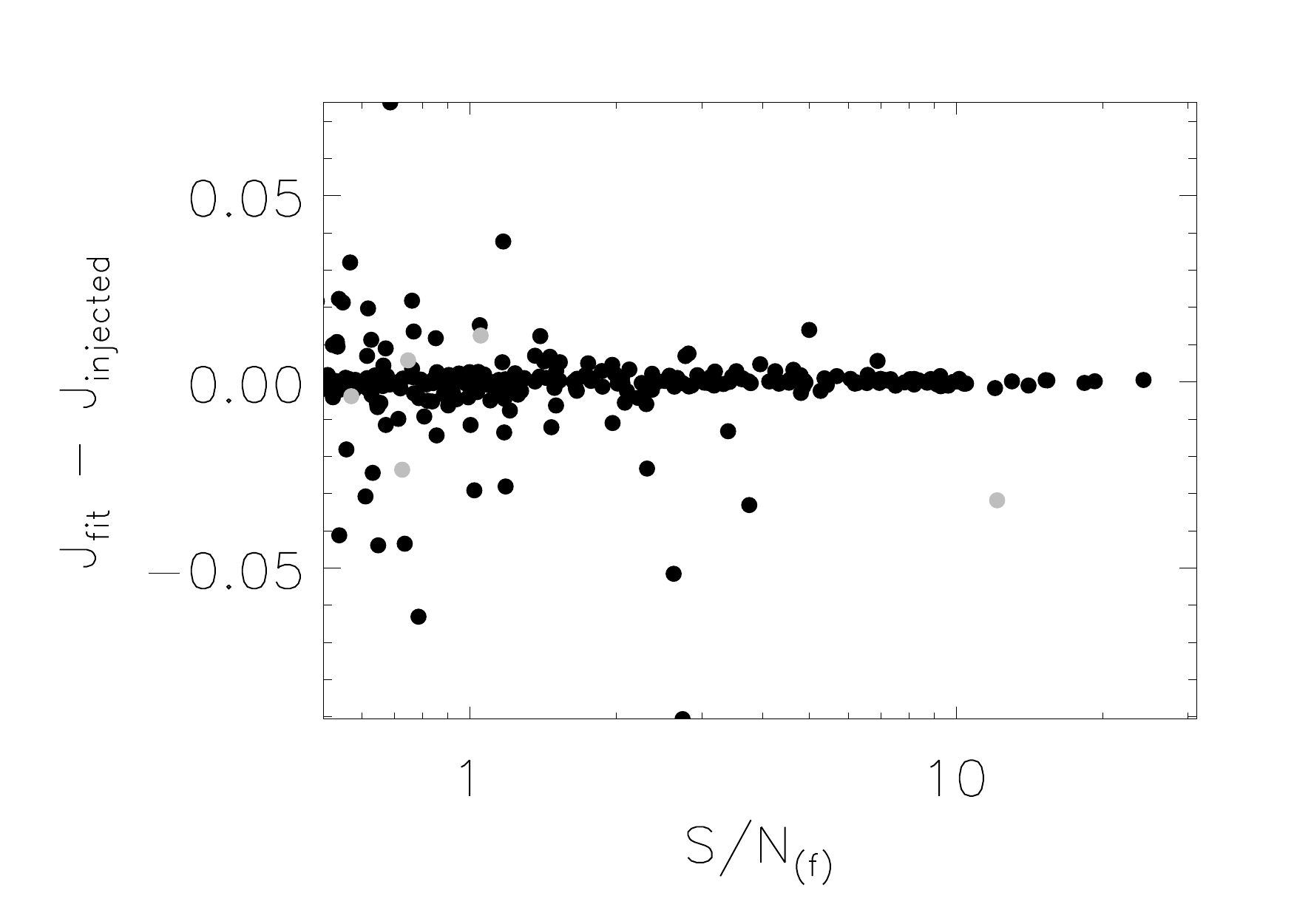}
      \caption{The result of the light curve test. The ordinate is the  S/N-ratio defined by Eq.~(\ref{eq:sn_surface_brightness}). The abcissa is the  difference between the simulated and the retrieved planet-to-star surface brightness ratio values. Note that in this Figure the x-axis has a logarithmic scale for better visibility. See the meaning of black/gray points at Figure~\ref{fig:prove_aRstar}. \label{fig:prove_surface_brightness}
      }
   \end{figure}

\begin{figure}
   \centering
   \includegraphics[width=9cm]{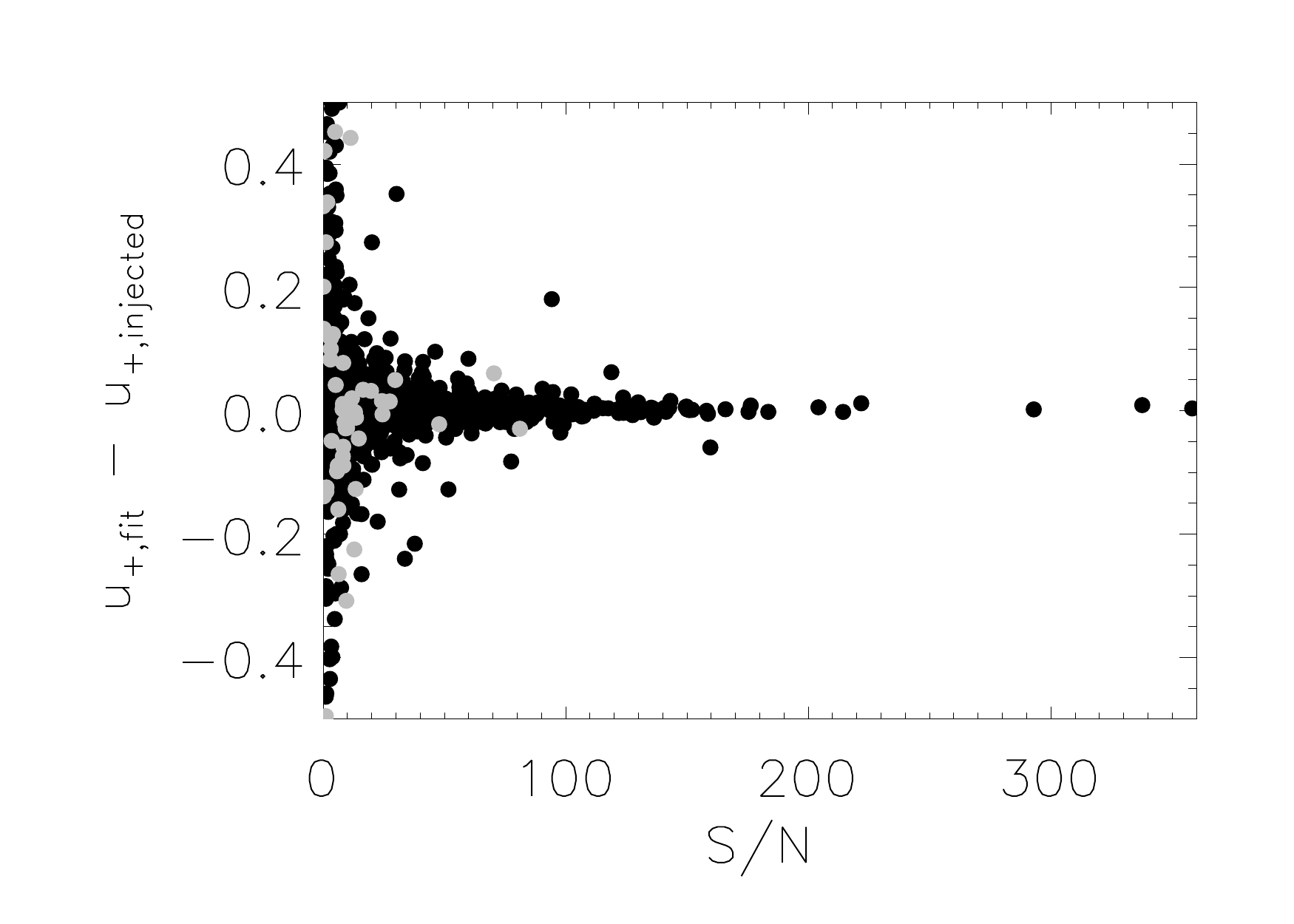}
      \caption{The result of the light curve test. The ordinate is the  S/N-ratio defined by Eq.~(\ref{eq:sn_transit}). The abcissa is the  difference between the simulated and the retrieved $u_+$ limb darkening coefficient combination values. See the meaning of black/gray points at Figure~\ref{fig:prove_aRstar}. \label{fig:prove_uplus}
      }
   \end{figure}

\begin{figure}
   \centering
   \includegraphics[width=9cm]{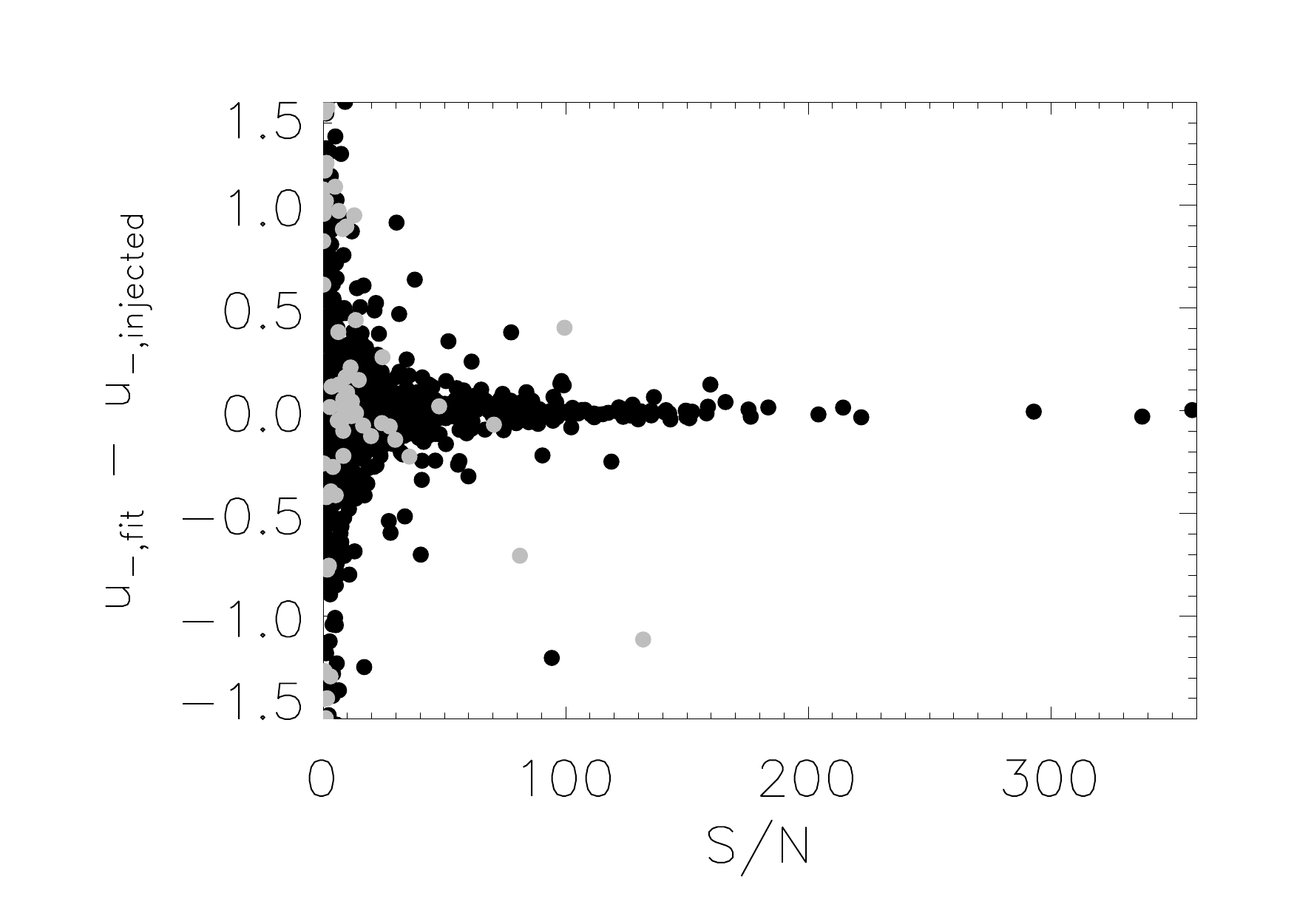}
      \caption{The result of the light curve test. The ordinate is the  S/N-ratio defined by Eq.~(\ref{eq:sn_transit}). The abcissa is the  difference between the simulated and the retrieved $u_-$ limb darkening coefficient combination values. See the meaning of black/gray points at Figure~\ref{fig:prove_aRstar}. \label{fig:prove_uminus}
      }
   \end{figure}

We also show the results of the tests in Figures~\ref{fig:prove_aRstar}-\ref{fig:prove_uminus}. From these figures we can read the minimum S/N-ratio needed to get reasonable accuracy in the parameter retrieval. We draw the following conclusions. When stellar variability or instrumental effects are present and they produce red noise in the light curve, we can set the following signal-to-noise ratio limits for the retrieval of the parameters with a wavelet+model fit with the following reasonable accuracies:

\begin{itemize}

\item[] $a/R_\mathrm{star}$: even at $S/N \sim 1$ we can get good results (better than 6\% relative error) and if $S/N > 3$ then we can get 2\% or better accuracy in the scaled semi-major axis ratio. This is not surprising because we set a Gaussian prior on the stellar radius which is strongly related to this parameter (via transit duration). We can safely assume that in most of the cases we know the stellar radius a priori from SED-fit combined with Gaia-parallax, asteroseismology or from other methods (e.g. Csizmadia 2021, under review at Astronomical Journal) (Figure~\ref{fig:prove_aRstar}).

\item[] $b$: if $S/N > 40$ the impact parameter can be retrieved with high accuracy. The impact parameter determination needs a precisely known stellar radius (3\% or better). If the stellar radius is less known (3-6\%) then most of the solutions lie in a good range, but some outliers appears (gray dots in Figure~\ref{fig:prove_b}). However, if we translate the impact parameter to inclination via $\cos i = b / (a/R_\mathrm{star})$, we find that the inclination values are always better than $\pm5^\circ$ when $S/N>40$ (Figure~\ref{fig:prove_inclination}). This causes very little difference in the planetary mass when the inclination value is used in the mass-function to determine the planetary mass. 

\item[] $R_\mathrm{planet}/R_\mathrm{star}$: the planet-to-star radius ratio is always better determined than $\pm2$\% if the $S/N > 50$. Even in the range of $10<S/N<50$ is better than 5\%. While \cite{morris2020} found that this precision cannot be reached with PLATO, our work differs from their work at two points. First, we did not take the effect of granulation into account but \cite{morris2020} did. We note that if the number of transits are small - in our example it was varied between 6 and 24 - then the granulation is not averaged out but it acts like pseudo-red noise \citep{chiavassa2017}.  \cite{morris2020} fitted their simulated light curves with Gaussian process and the residuals was still found in the order of 100 ppm. We leave the issue open whether the wavelets can manage the effect of granulation but there is a possibility for that. Second, we considered that the stellar radius is known by at least 2\% accuracy as a SED-fit or future  asteroseismological part of PLATO will provide this for its primary sample and we used this prior in our fit - they did not. Then, if the granulation is negligible or it can be averaged out by many transit measurements being a white noise, the use of the asteroseismological or SED-fit based stellar radius constraints in the fit are able to provide radius ratio values what PLATO needs even for a Sun-Earth radius ratio ($k \sim 0.009$) (\citealt{rauer2014}, Figure~\ref{fig:prove_radiusratio}).

\item[] $q$: When $S/N_{(q)} > 20$ then the approach is able to recover the mass ratio with a better accuracy than 10\% with some rare exceptions when the reached accuracy is just 20\%. This is enough to validate a planet candidate and it may confirm the planetary mass measurement independently of RV  (Figure~\ref{fig:prove_massratio}).

\item[] $A_\mathrm{planet}$: The geometric albedo of the planet can be retrieved with at least $\pm0.05$ accuracy if  $S/N_{(A)}>11$( Figure~\ref{fig:prove_albedo2}).The accuracy increases fast as the signal-to-noise ratio increases.

\item[] $\varepsilon$: To get the the value of the reflection shift with this wavelet-based filtering method with a $\pm4^\circ$ accuracy one needs at least $S/N_{(A)}=11$ while to get it with better accuracy than $\pm2^\circ$ $S/N_{(A)}<25$ is needed (Figure~\ref{fig:prove_reflectionshift}).

\item[] $J$: To recover the surface brightness ratio of the star and the planet - which is possible form occultations - one needs $S/N_{(J)}>10$ (Figure~\ref{fig:prove_surface_brightness}).

\item[] limb darkening: The limb darkening coefficient combinations $u_+$ and $u_-$ can be retrieved with $\pm0.01$ accuracy of $S/N > 100$ but in some rare cases exceptions occur. (Figures~\ref{fig:prove_uplus} and ~\ref{fig:prove_uminus}). 

\end{itemize}

We add that these results cannot be reached if we do not have a good prior on the stellar radius which helps to constrain the transit duration and thus the impact parameter.

We make the note that, of course, wavelets cannot replace the variable contamination effect. If the contamination is variable from sector to sector of TESS or variable from frame to frame as a consequence of the rotation of CHEOPS, for instance, then this contamination must be corrected before fitting procedure or it must be taken into account in the system's model and not in the wavelet-model.

%
%

%
%

%
%

%
%

%
%

%
%

%
%

%
%

%

%
   
%

\section{Summary \& Conclusions} \label{sec:summary}

We have shown by numerical tests and injecting planetary light curve models that in an ideal case, where the right model of physical reality is well known, the wavelets are able to reconstruct and to filter out the stellar variability and instrumental noise effects, like jumps, cosmic ray hits, discontinuities, detector ageing etc. In Section~\ref{sec:wavelet_method} we have given limits in terms of signal-to-noise ratio for the accuracy of the planet and system parameter retrieval. The wavelet-approach worked well on a wide variety of possible noise sources and stellar variability phenomena, and it was able to manage even high amplitude or fast variable stellar variability and instrumental noise sources (see the examples in Figures~\ref{fig:jump_example}-\ref{fig:pulsation2_example} 
).

To reach this high performance, we needed a penalty function during the optimization and uncertainty estimation process. The penalty function decreased the likelihood of the solution if the root mean square of the residuals of the system+wavelet fit deviated from the average uncertainty of the data points. In other words, we prescribed the white noise level what the wavelet-based noise modelling must reach. Without such a precondition, there is a danger of overfit, i.e. we fit everything with combinations of wavelets instead of determining the noise properties of the light curve and to extract the system information. This is further illustrated by the example of KELT-9b stellar pulsational like variability in Paper II: it was fully modelled with wavelets and the pulsation was not visible in the red noise corrected flux-residuals. This example in Paper II sets a caveat: every unmodelled, unkown effect will be incorporated into the  wavelets and the information is lost. Therefore, a good model must be selected for the fits when one works with wavelet-based noise-models.

However, data overfit can be done with other methods as well, for example with Gaussian processes. In addition, the wavelet procedure of \cite{carterwinn09} used here, needs only two free parameters. In Gaussian processes, the number of free parameters can be much larger and it can be that one selects a non-appropriate kernel for that noise modelling approach. While the red noise factor of the wavelet-based noise model has no any physical meaning, sometimes the Gaussian process parameters can be linked to some physical process.

We left the limb darkening coefficients free in the test. One can imagine that applying a good prior on the limb darkening may further increase the performance and we can get better results at even lower signal-to-noise ratios. See \citep{csizmadia11} how impact parameter, scaled semi-major axis, planet-to-star radius ratio is degenerated with limb darkening coefficients. However, the present knowledge of limb darkening prefers to leave the limb darkening coefficients free parameters in the fit \citep{csizmadia13, espinoza15, agol2020}.

We also validated the gravity darkening treatment of TLCM for planets with modelling the Kepler light curve of Kepler-13Ab, a well-known object with asymmetric transits. We found results which are fully compatible with the spin-orbit angle $\lambda$ obtained by Doppler-tomography results \citep[][]{johnson14} and with the stellar inclination value of \cite{szabo20} within 2 degrees which is within the error bars. 

The latest version of TLCM with the updated ellipsoidal and reflection effects will be available at \url{http://www.transits.hu} once this paper is accepted.

\begin{acknowledgements}
       The authors gratefully acknowledge the European Space Agency and the PLATO Mission Consortium, whose outstanding efforts have made these results possible.

       We thank DFG Research Unit 2440: ’Matter Under Planetary Interior Conditions: High Pressure, Planetary, and Plasma Physics’ for support. We also acknowledge support by DFG grants RA 714/14-1 within the DFG Schwerpunkt SPP 1992: ’Exploring the Diversity of Extrasolar Planets’. CsSz also thanks the Hungarian National Research, Development and Innovation Office for the NKFIH -- OTKA KH-130372 grant.
\end{acknowledgements}



\bibliographystyle{aa}
\bibliography{pow.bib}

\end{document}